\documentclass[preprintnumbers,amsmath,11pt,amssymb,floatfix,superscriptaddress,nofootinbib]{article}

\def\mytitle#1{\setcounter{equation}{0}
\setcounter{footnote}{0}
\begin{flushleft}\Large\textbf{#1}\end{flushleft}
\vspace{0.25cm}}
\def\myname#1{\leftline{{\large #1}}\vspace{-0.13cm}}
\def\myplace#1#2{\small\begin{flushleft}\textit{#1}\\
\texttt{#2}\end{flushleft}}

\def\myclassification#1{\small\noindent
Keywords :
       #1\vspace{0.5cm}}
\usepackage{graphicx}
\usepackage{amsmath}
\begin{document}
\mytitle{Geometrothermodynamic Analysis and $P$-$V$ criticality of Higher Dimensional Charged Gauss-Bonnet Black Holes With First Order Entropy Correction}

\myname{$Amritendu~ Haldar^{*}$\footnote{amritendu.h@gmail.com} and $Ritabrata~
Biswas^{\dag}$\footnote{biswas.ritabrata@gmail.com}}
\myplace{*Department of Physics, Sripat Singh College, City: Jiaganj-742123, Dist.: Murshidabad , State: West Bengal, India.\\$\dag$ Department of Mathematics, The University of Burdwan, Golapbag Academic Complex, City: Burdwan-713104, Dist.: Purba Burdwan, State: West Bengal, India.} {}
 
\begin{abstract}
We consider a charged Gauss-Bonnet black hole in $d$-dimensional spacetime and examine the effect of thermal fluctuations on the thermodynamics of the concerned black hole. At first we take the first order logarithmic correction term in entropy and compute the thermodynamic potentials like Helmholtz free energy $F$, enthalpy $H$ and Gibbs free energy $G$ in the spherical, Ricci flat and hyperbolic topology of the black hole horizon, respectively. We also investigate the $P$-$V$ criticality and calculate the critical volume $V_c$, critical pressure $P_c$ and critical temperature $T_c$ using different equations when $P$-$V$ criticality appears. We show that there is no critical point without thermal fluctuations for this type of black hole. We find that the presence of logarithmic correction in it is necessary to have critical points and stable phases. Moreover, we study the stability  of the black holes by employing the specific heat. Finally, we study the geometrothermodynamics  and analyse the Ricci scalar of the Ruppeiner metric graphically for the same.
\end{abstract}

\myclassification{Thermal fluctuations; Thermodynamic potentials; Ricci flat;  Ruppeiner metric.}

\section{Introduction:}
General relativity predicts local singularities like black holes (BHs hereafter) which are fascinating topics of research in present time. A BH itself has a singularity at the origin of the spacetime where the curvatures, density etc. become infinite \cite{Hawking 1973, Hawking 1996}. Again from a BH no mattaer, energy and electromagnetic waves such as light can escape. This singularity at $r=0$ is assumed to be wrapped by another singularity which is named as event horizon, crossing which from outside to inside, the role of time and space coordinates swap each other's role. Applying quantum field theoty, Hawking  \cite{Hawking 1975} has shown that the BHs have a thermal radiation which provides a real connection between quantum mechanics and gravity. Another interesting fact of BHs is that the laws of BHs on the event horizon $(r=r_+)$ resemble to the laws of classical thermodynamics. This exhibits that each BH may be treated as thermodynamic object. Therefore, it is very much significant to study the thermodynamics of BHs. In the references \cite{Bekenstein 1972, Bekenstein 1973, Bekenstein 1974, Bardeen 1973, Hawking 1975, Wald 2001, Page 2005}, the authors have studied the thermodynamics of different kinds of BHs. The thermodynamic properties of BHs in extended phase have also been studied in the references\cite{Hawking 1983, Peca 1999} which show, in a particular phase that the thermodynamic parameters behave smoothly. However, some of parameters such as specific heat among them manifest a discontinuity as the BH's temperature varies. The Hawking-Page phase transition may be explained as the confinement/deconfinement phase transition of gauge field in the AdS/CFT (Conformal Field Theory) correspondence \cite{Witten 1998}. In Anti-de Sitter (AdS hereafter) space and in de Sitter space, the thermodynamics of BHs are not exactly the same. The large BHs are thermodynamically stable and have positive heat capacity in AdS space, whereas, the small BHs are thermodynamically unstable and as they have negative heat capacity, become hotter and eventually evaporate.   

The holographic principle \cite{Susskind 1995, Bousso 2002} which is inspired by aforementioned entropy equation and the modified entropy is due to quantum gravity corrections depicted latter equates the degrees of freedom (DoF hereafter) of the boundary to any other region of space. This principle will be corrected near the Planck scale and the quantum gravity corrections modify the topology of space-time at this scale \cite{Rama 1999, Bak 2000}. The entropy $S_+$ of BHs is related to the surface area $A$ of the BHs at event horizon (as both entropy and surface area are ever increasing parameters) as: $S_+=\frac{A}{4}$, where $ A=\int ^{2\pi}_{0} \int ^{\pi}_{0} \sqrt{g_{\theta \theta}(r_+){g_{\phi\phi}(r_+)}} {d\theta d\phi} = 4\pi r_+^2.$ The corrected entropy  may be expressed as $S=S_0+ \alpha ln A+ \gamma_1 A^{-1}+ \gamma_2 A^{-2}+........ $, where $\alpha, \gamma_1, \gamma_2,......$ are the coefficients which depend on different BH thermodynamic parameters \cite{Upadhyay 2017}. Moreover, the area dependance has been computed by specific model such as G$\ddot{o}$del like BH \cite{Pourdarvish 2013} where the corrected thermodynamics has heen studied by utilizing the modified form of entropy $S=S_0+ \alpha ln A$. Further,  one can investigate the corrected thermodynamics of BHs by employing the non-perturbative quantum gravity. With the help of Cardy formula, the effects of quantum correction to the BH thermodynamics have been investigated in the reference\cite{Govindarajan 2001}. The corrected thermodynamics of BHs have been studied under the effect of matter field around the BHs \cite{Mann 1998, Medved 1999, Medved 2001}. In many literatures \cite{Solodukhin 1998, Sen 2011, Sen 2013, Lowe 2010}, the thermodynamic corrections produced by string theory which are in agreement with the other approaches to quantum gravity have been studied. The corrected thermodynamics of a dilatonic BH has also been discussed and observed to have the same universal manner \cite{Jing 2001}. The partition function of a BH contributes a significant role to study the corrected thermodynamics of a BH \cite{Birmingham 2001}. The generalized uncertainty principle (GUP hereafter) helps us to yield the logarithmic correction \cite{Ali 2012, Faizal1 2015} and is also very much useful to study the corrected thermodynamics of a BH in agreement with the other which approaches to quantum gravity. The thermal fluctuations in the  BH thermodynamics would be obtained from a quantum correction to the space-time topology and it is of the same form that is as expected from the quantum gravitational effects \cite{Das 2002,  More 2005, Sadeghi 2014}.

One can investigate the natures of phase transitions  of thermodynamic systems more perfectly by the concept of geometry of thermodynamics. In this method, the thermodynamic line elements and curvatures can be interpreted as a system interaction.
One can study the phase transitions of thermodynamic systems by obtaining the curvature singularities of the thermodynamic metric. The curvature singularities of the thermodynamic metric (calculated in GTD approach) for AdS space have been studied in the reference \cite{Quevedo2008}. At first, Weinhold defined the second derivatives of the internal energy $U$ connection with entropy $S$ and the electric charge $Q$ or any other thermodynamic parameter of a thermodynamical system to introduce a Riemannian metric \cite{Weinhold1, Weinhold2} as:
\begin{equation}\label{ah7_equn_1}
g^W=\frac{\partial^2 M}{\partial X^i\partial X^j}dX^idX^j,~~ X^i=X^i(S,Q).
\end{equation}
Another metric which is introduced by Ruppeiner \cite{Ruppeiner1, Ruppeiner2} is defined as the negative Hessian of the entropy with respect to the internal energy and other extensive quantities of a thermodynamic system and is expressed as:
\begin{equation}\label{ah7_equn_2}
g^W=\frac{\partial^2 S}{\partial Y^i\partial Y^j}dY^idY^j,~~ Y^i=Y^i(M,Q)~~.
\end{equation}
These two metrics are related as:
\begin{equation}\label{ah7_equn_3} 
ds^2_{R}=\frac{1}{T}ds^2_{W}~~.
\end{equation}  
Divergence in the Ricci scalar of Ruppeiner metric can speculate about the phase transitions.
It was shown that the phase transition points of the heat capacity do not match those in the Weinhold and Ruppeiner metrics at present.
In \cite{Pourhassan1}, the authers have studied the effects of quantum correction on the BH thermodynamics for the LMP solution of Horava Lifshitz BH in flat, spherical and hyperbolic spaces. Quantum corrections to thermodynamics of quasitopological BHs are studied by Upadhyay, S. \cite{Upadhyay1}. It is found from this that there is a critical horizon radius for total mass density. Thermodynamics of a BH geometry with hyperscaling violation is studied in the reference \cite{Pourhassan2}. Effects of thermal fluctuations on the thermodynamics of massive gravity BHs in four dimensional AdS space have been studied in reference \cite{Upadhyay2}. 

There are several articles in which the quantum corrections have already been used to study the BH geometries. High curvature BTZ BHs' heat capacity, free energy and the geometric thermodynamics have been studied in reference \cite{Hendi1}. {\bf In the reference \cite{Nadeem}, the authors have discussed the thermal fluctions of the thermodynamics of small non-rotating BTZ BH. The  thermal fluctions of charged BHs in gravity's rainbow have been studied in reference\cite{Panahiyna 2018}}. The authors in the reference \cite{Pourhassan3} have studied the termodynamics of higher order entropy corrected Schwarzschild Beltrami-deSitter BH. The stability of Van der Waals' BHs in presence of logarithmic correction has been studied in \cite{Upadhyay3}. Thermodynamic geometry of a static BH in $f(R)$ gravity has been studied in \cite{Upadhyay4}. The authors have studied the logarithamic correction of the entropy of an AdS charged BH in \cite{Pourhassan 2015} and found that the thermodynamics of the AdS BH is modified due to the thermal fluctuations. While the effects of thermal fluctuations on the thermodynamics of a modified Hayward BH have been studied, it is noticed that the thermal fluctuations reduce the internal energy $U$ and pressure $P$ of that BH \cite{Pourhassan1 2016}. The thermal fluctuations for a black saturn  as well as a charged dilatonic black saturn have been investigated in the letaratures \cite{Faizal2 2015} and \cite{Pourhassan2 2016}. For the black saturn, it has been found that the thermal fluctuations do not have the major effects on stability of BHs. After investigating the thermodynamic properties of a small spinning Kerr-AdS BH under the effects of thermal fluctuations \cite{Pourhassan3 2016}, one can conclude that the logarithmic correction in entropy plays an important role for a sufficiently small BH. The authors have studied the logarithmically corrected thermodynamics of a dyonic charged AdS BH in \cite{Sadeghi 2016}, which is holographic dual of a Van der Waals' fluid. {\bf With the consideration of a charged rotating AdS BH in four dimensions, the effects of leading-order corrections on the thermodynamics of such system have been studied in \cite{Upadhyay 2018}.}        

Our objective in this paper is to investigate the effect of thermal fluctuations on the thermodynamics of the charged Gauss-Bonnet BHs in $d$-dimensional space-time. Initially, we will take the first order logarithmic correction term in entropy and find the thermodynamic potentials like Helmholtz free energy $F$, enthalpy $H$ and Gibbs free energy $G$ and also the internal energy $U$ in the spherical, Ricci flat and hyperbolic topology of the black hole horizon, respectively and try to interpret them graphically . We will also investigate the $P$-$V$ criticality and calculate the critical volume $V_c$, critical pressure $P_c$ and critical temperature $T_c$ using different equations. We will find that there is no critical point without thermal fluctuations for this BH. We will also find that presence of logarithmic correction in it is necessary to have critical points and stable phases. Here we will also calculate the ratio $\frac{P_c r_c}{T_c}$ and find that this ratio is a universal number and only depends on the electric charge $Q$ of the BH. Moreover, we will study the stability  of the BHs by employing the specific heat. Finally, we will study the geometrothermodynamics  and analyses the Ricci scalar of the Ruppeiner metric graphically for the same.

This paper is organized as follows: in the next section, we will present the brief thermodynamics of $d$-dimensional charged Gauss-Bonnet BHs. We will consider the first order correction to entropy (i.e., the logarithmic corrected entropy) as leading order of thermal fluctuation. In section 3, we will analyze the $P-V$ criticality and examine the stability of the BH by employing the specific heat of BHs in section 4. We will study the geometrothermodynamics for the thermodynamic system in section 5. Finally, we will present conclusion of the work in the last section.            
\section{$d$-Dimensional Charged Gauss-Bonnet Black Holes:}

The action for Einstein-Gauss-Bonnet-Maxwell model may be written with negative cosmological constant $\Lambda$ as (Here we have used the Planck units, i.e., $G=c=k_B=\hbar=1$ ) \cite{Cai 2013}

\begin{equation}\label{ah7_equn4}
{\cal A}=\frac{1}{16\pi}\int d^dx\sqrt{-g}\left[R-2\Lambda+\alpha_{GB}\left(R^2-4R_{ab}R^{ab}+R_{abcd}R^{abcd}-4\pi{\cal F}_{ab}F^{ab}\right)\right],
\end{equation}
where $\alpha_{GB}$ is the Gauss-Bonnet coefficient having dimensions of $[length]^2$ and is positive in the
heterotic string theory \cite{Kubiznak 2012}. Therefore, in our works we restrict in $\alpha_{GB}>0$ only. ${\cal F}_{ab}$ is the Maxwell's field strength, defined as ${\cal F}_{ab} = {\partial}_{a} A_{a} − {\partial}_{b} A_{b}$ with vector potential $A_ {a}$. As the Gauss-Bonnet term $\left(=R^2-4R_{ab}R^{ab}+R_{abcd}R^{abcd}\right)$ is a topological invariant in four dimensions. So $d \geq 5$ is considered in this paper.

We assume the metric being of the following form
\begin{equation}\label{ah7_equn5}
ds^2=-f(r)dt^2+\frac{dr^2}{f(r)}+r^2h_{ij}dx^i dx^j,
\end{equation}
where $h_{ij}dx^i dx^j$ denotes the line element of a $(d - 2)$ -dimensional maximal symmetric Einstein space with constant curvature $(d - 2)(d - 3)$ $k$ and volume $\Sigma_k$. Without loss of the generality, one may take $k = 1, 0$ and $-1$, corresponding to the spherical, Ricci flat and hyperbolic topology of the BH horizon, respectively. The metric function $f(r$) is given by the references\cite{LIng 2007, Li 2009, Ali 2014, Kim 2016, Feng 2017}
\begin{equation}\label{ah7_equn6}
f(r)=k+\frac{r^2}{2 \tilde{\alpha}}\left(1-\sqrt{1+\frac{64 \pi \tilde{\alpha} M}{(d-2)\Sigma_k r^{d-1}}+\frac{2 \tilde{\alpha} Q^2}{(d-2)(d-3)r^{2d-4}}+\frac{8 \tilde{\alpha}\Lambda}{(d-1)(d-2)}}\right),
\end{equation}
where $\tilde{\alpha} = (d - 3) (d - 4)\alpha_{GB}$, $M$ and $Q$ are the mass and charge of the BHs and $P$ $ = -\frac{\Lambda}{8 \pi}= \frac{(d-1)(d-2)}{16 \pi l^2}$, here $l$ signifies the AdS radius of BHs. In order to have a well-defined vacuum solution with $M = Q = 0$, the effective Gauss-Bonnet coefficient $\tilde{\alpha}$ and pressure $P$ have to satisfy the following constraint given as:
\begin{equation}\label{ah7_equn7}
0< \frac{64 \pi \tilde{\alpha} P}{(d-1)(d-2)}\leq 1~~.
\end{equation}
On the event horizon $r = r_+$, mass of this BH can be expressed as 
\begin{equation}\label{ah7_equn8}
M=\frac{ 2 (d-3) \Sigma_k r_+^{2 d} \left\{(d-2) (d-1) k \left(k \tilde{\alpha} +r_+^2\right)-2 r_+^4 \Lambda \right\}+(1-d) Q^2 r_+^8 \Sigma_k }{32(d-3) (d-1) \pi r_+^{d+5}}~~.
\end{equation}
Correspondingly, the temperature of the BH on the horizon could be expressed as
$$T_+=\frac{f^{'}(r_+)}{4\pi}=\frac{r_+^{-2 d-1}}{8  \tilde{\alpha}  \pi (d-2)\left(2 k \tilde{\alpha} +r_+^2\right)}\bigg[2 (d-5) (d-2) k^2 \tilde{\alpha} ^2 r_+^{2 d}+2 (d-5) (d-2) k \tilde{\alpha}  r_+^{2 d+2}$$
\begin{equation}\label{ah7_equn9}
+2 r_+^{2 d+4}\left\{\frac{2 k \tilde{\alpha}}{r_+^2}(d-2) -2\tilde{\alpha}  \Lambda \right\}+Q^2r_+^8 \tilde{\alpha} \bigg]~~.
\end{equation}
\begin{figure}[h!]
\begin{center}
~~~~~~~~~~~~~~~~~~Fig.-1.1a ~~~~~~~~~~~~~~~Fig.-1.1b~~~~~~~~~~~~~~~~1.1c~~~~~~~~~~~~~~~~\\
\includegraphics[scale=.5]{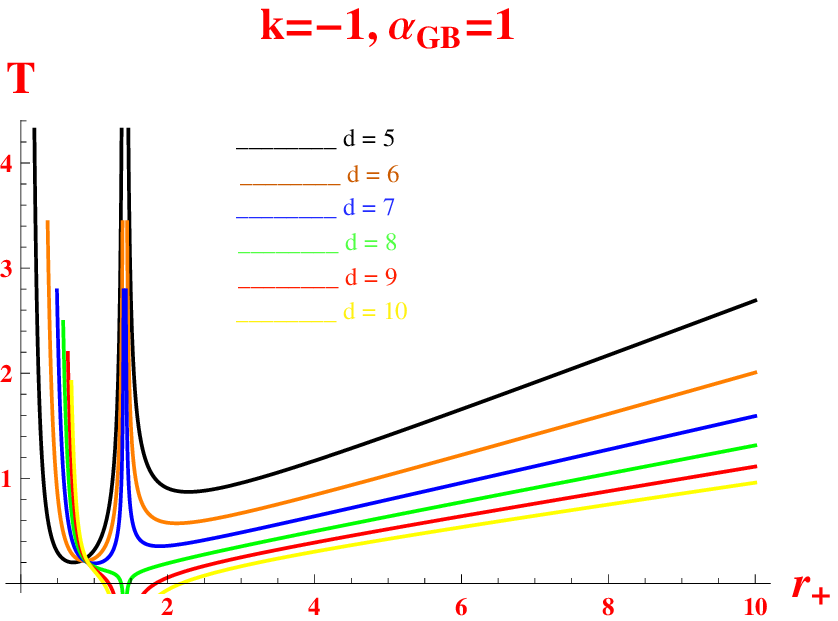}
\includegraphics[scale=.5]{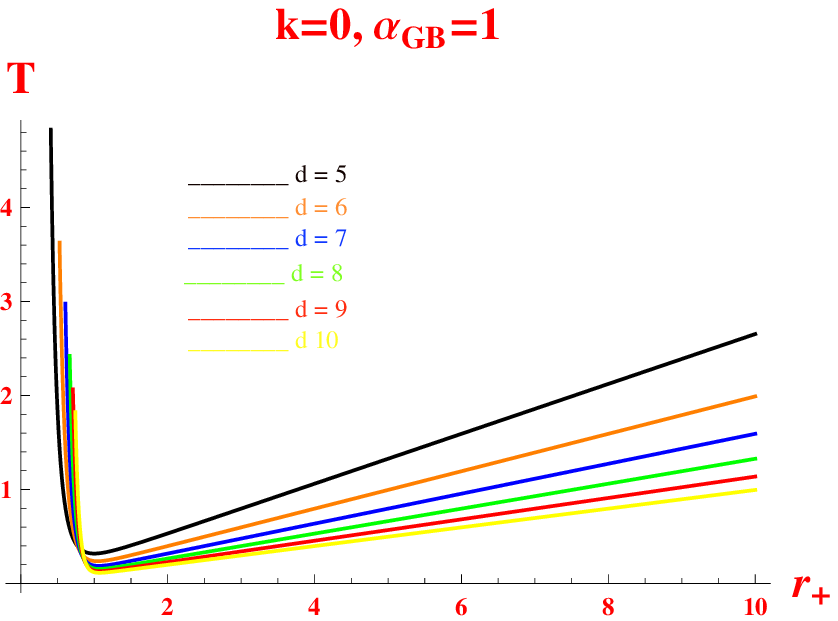}
\includegraphics[scale=.5]{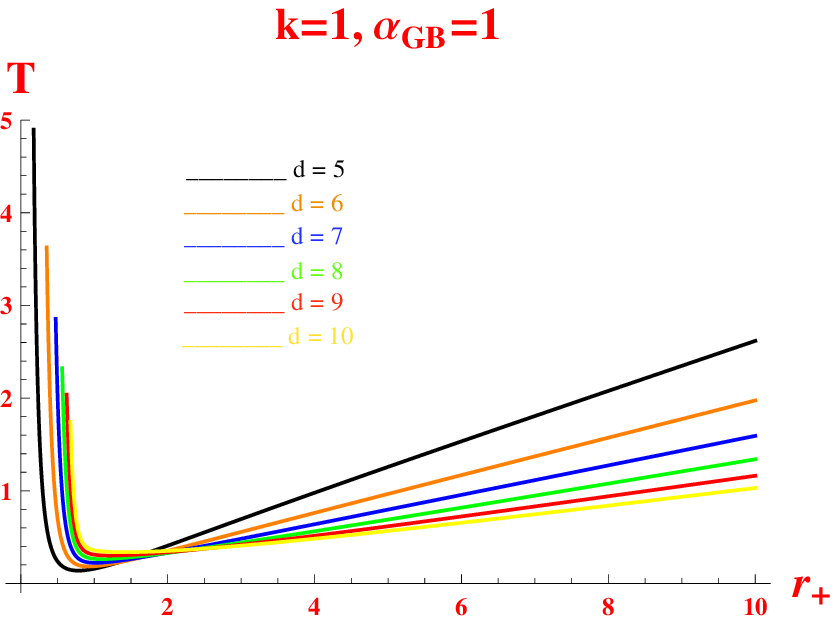}\\
~~~~~~~~~~~~~~~~~~Fig.-1.2a ~~~~~~~~~~~~~~~Fig.-1.2b~~~~~~~~~~~~~~~~1.2c~~~~~~~~~~~~~~~~\\
\includegraphics[scale=.5]{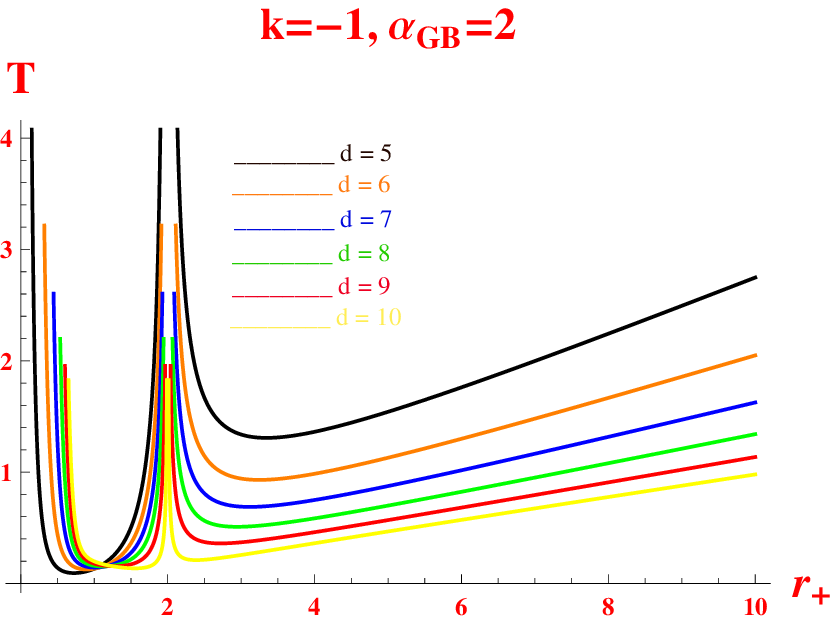}
\includegraphics[scale=.5]{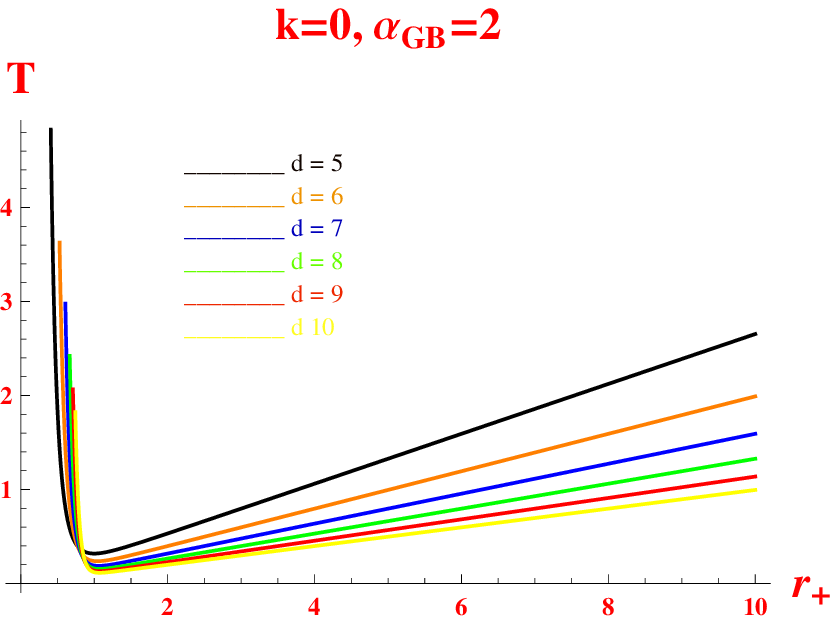}
\includegraphics[scale=.5]{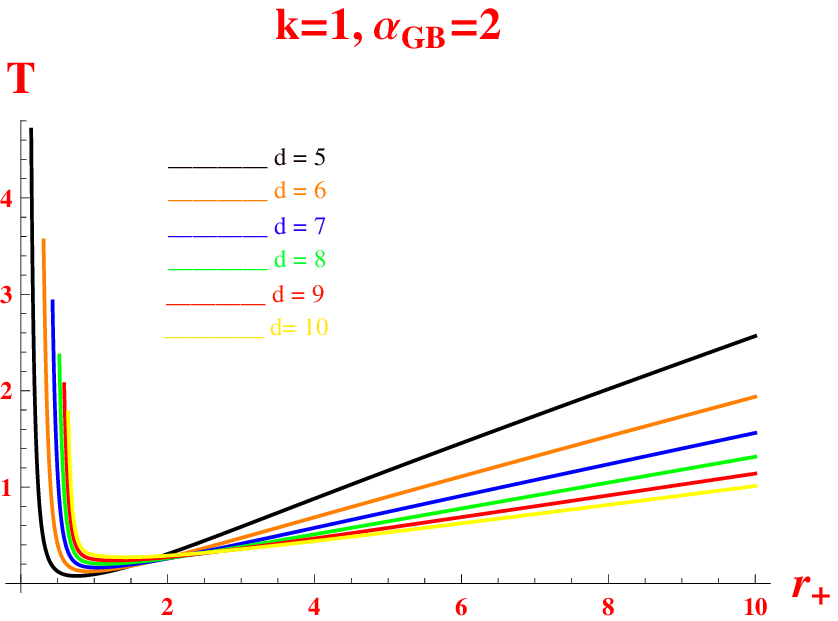}\\
Fig.-1.1a-1.1c represent the variation of $ T_+ $ with respect to horizon radius $ r_+ $ for $ k=-1, 0, 1 $, $\alpha_{GB}=1$ and different dimensions $d$. \\
Fig.-1.2a-1.2c represent the variation of $ T_+ $ with respect to horizon radius $ r_+ $ for $ k=-1, 0, 1 $, $\alpha_{GB}=2$ and different dimensions $d$. \\
\end{center} 
\end{figure}
Fig.-1.1a-c depict the variations of $ T_+ $ with respect to horizon radius $ r_+ $ for $ k=-1, 0, 1 $, $\alpha_{GB}=1$ and for different dimensions $d$. Here we observe that for $k=-1$, a sharp maximum is found for $d=5, 6$ and $7$ whereas for $d=8, 9$ and $10$ a sharp minimum is observed at low horizon region. These minimum values go to negative temperature region which imply that for very high dimensions, BHs are unphysical. But at high horizon region for all dimensions temperatures are positive and increasing functions of $r_+$ and this signifies that for very high dimensions, the large BHs have their physical existence. For $k=0$, we find at low horizon region, initially $T_+$ decreases steeply due to slight increment of $r_+$ and further increment of $r_+$, $T_+$ increases gradually. The rate of increment decreases due to increment of $d$. For $k=1$, we get the curve as almost similar nature as for $k=0$ case.
Fig.-1.2a-c depict the same for $\alpha_{GB}=2$. Here we find that for $k=-1$, a sharp maximum for all dimensions at low horizon region exist. For $k=0$ and $1$ we have the curves as similar as for $\alpha_{GB}=1$ case.
      
In the reference \cite{Sommerfield 1956}, the authors show that the first order correction to entropy of BH is proportional to $\ln\left(C_V T^2\right)$, where $C_V=\frac{1}{T^2}\left(\frac{\partial^2 S}{\partial \beta^2}\right)_{\beta=\beta_0}$ . Following some literatures as \cite{Pradhan 2016, Das 2001} we can calculate that $\frac{1}{T^2}\left(\frac{\partial^2 S}{\partial \beta^2}\right)_{\beta=\beta_0}=S_0$. So the correction term might be written to be proportional to $\ln \left(S_0T^2\right)$ and hence, the first order corrected form of the entropy may be written as:
\begin{equation}\label{ah7_equn10}
 S=S_0-\frac{1}{2}\ln(S_0T_+^2)~~,
\end{equation}
where $S_0$ is the zeroth entropy which is experessed in $d$-dimensional space-time as \cite{XU 2015}
\begin{equation}\label{ah7_equn11}
S_0= \pi r_+^{(d-2)}~~.
\end{equation}
Using the $equations$ (\ref{ah7_equn9}), (\ref{ah7_equn10}) and (\ref{ah7_equn11}) we obtain the corrected entropy as:
$$S=\pi r_+^{(d-2)}-\frac{1}{2}\ln \bigg[\frac{r_+^{-4d}}{64(d-2)^2(2k\tilde{\alpha}+r_+^2)^2 \pi}\bigg(2 (d-5) (d-2) k^2 \tilde{\alpha} ^2 r_+^{2 d}+2 (d-5) (d-2) k \tilde{\alpha}  r_+^{2 d+2}$$
\begin{equation}\label{ah7_equn12}
+2 r_+^{2 d+4}\left\{\frac{2 k \tilde{\alpha}}{r_+^2}(d-2) -2\tilde{\alpha}  \Lambda \right\}+Q^2r_+^8 \tilde{\alpha} \bigg)^2\bigg].
\end{equation} 
\begin{figure}[h!]
\begin{center}
~~~~~~~~~~~~~~~~~~Fig.-2a ~~~~~~~~~~~~~~~Fig.-2b~~~~~~~~~~~Fig.-2c~~~~~~~~~~~\\
\includegraphics[scale=.5]{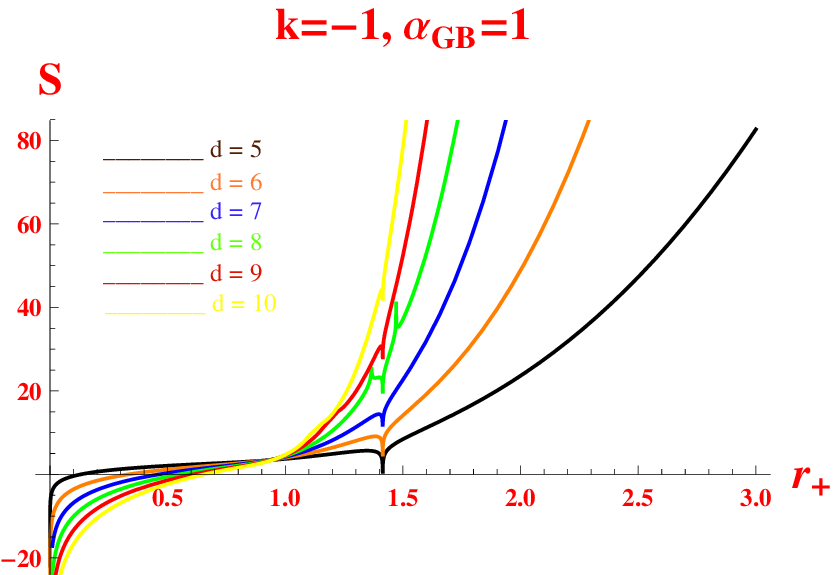}
\includegraphics[scale=.5]{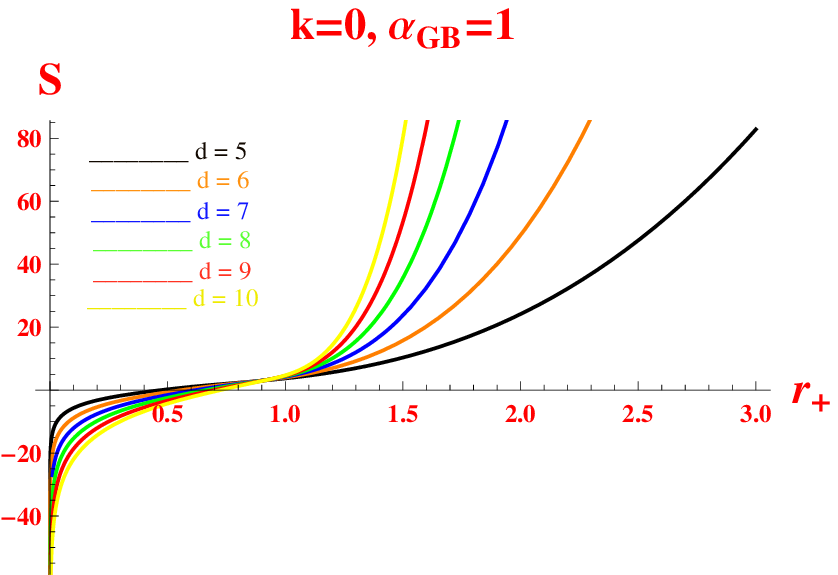}
\includegraphics[scale=.5]{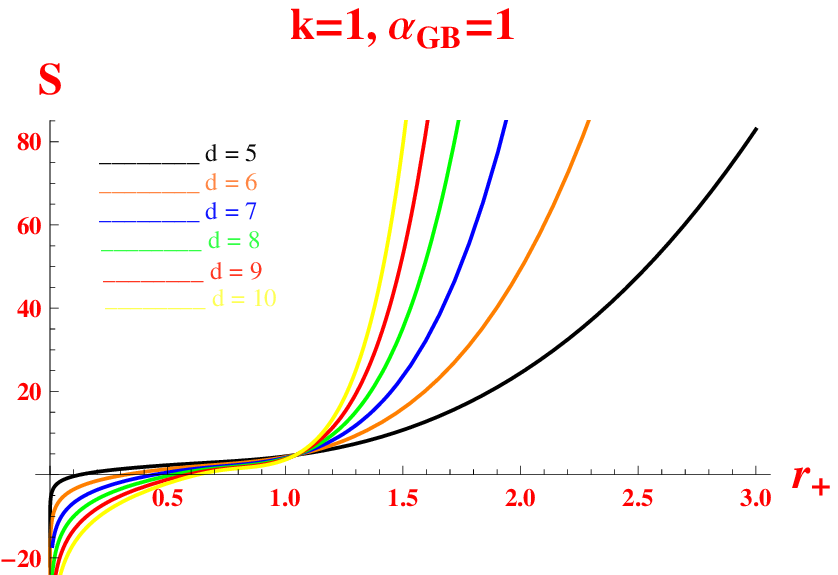}\\
Fig.-2a-2c represent the variation of $ S $ with respect to horizon radius $ r_+ $ for $ k=-1, 0, 1 $ and for different dimensions $d$.
\end{center} 
\end{figure}
The variation of BH entropy $S$ with respect to $r_+$ for $ k=-1, 0, 1 $ and for different dimensions $d$ are shown in the Fig. 2a-c. The average natures of the curves are same, i.e., for increase of $r_+$, $S$ is increased as expected. But here the inclinations of the curves are increased due to increase of $d$.     
    
The Helmholtz free energy $F$, related to entropy $S$ and temperature $T_+$ is given by,
\begin{equation}\label{ah7_equn113}
F=-\int {S dT_+}
\end{equation}
which gives
$$F=\frac{r^{-2 d-5}}{640 (d-2) k^{3/2} \alpha ^{5/2} \pi  \left(2 k \alpha +r^2\right)}$$
$$\bigg[2 \sqrt{k} \sqrt{\alpha } \bigg(d r^{2 d} \left(2 k \alpha +r^2\right) \left(-12 (d-5) (d-2) k^2 \alpha ^2-10 (d-2) (d-1) k r^2 \alpha\right.$$
$$+15 r^4 ((d-3) d+8 \alpha  \Lambda +2)\bigg)+20 k \alpha ^2 \left(2 r^{2 d} \left((d-2) k \left((d-5) k \alpha +(d-3) r^2\right)-2 r^4 \Lambda \right)\right.$$
$$+Q^2 r^8\big) \log \left(\frac{r^{-3 d-4} \left(2 r^{2 d} \left((d-2) k \big((d-5) k \alpha+(d-3) r^2\right)-2 r^4 \Lambda \right)+Q^2 r^8\big)^2}{64 (d-2)^2 \pi  \left(2 k \alpha +r^2\right)^2}\right)$$    
\begin{equation}\label{ah7_equn14}
 +15 \sqrt{2} d r^{2 d+5} ((d-3) d+8 \alpha  \Lambda +2) \left(2 k \alpha +r^2\right) \tan ^{-1}\left(\frac{r}{\sqrt{2} \sqrt{k} \sqrt{\alpha}}\right)\bigg]
\end{equation}
\begin{figure}[h!]
\begin{center}
~~~~~~~~~~~~~Fig.-3.1a ~~~~~~~~~~~~~~~Fig.-3.1b~~~~~~~~~~Fig.-3.2a ~~~~~~~~Fig.-3.2b~~~~~~~~~\\
\includegraphics[scale=.4]{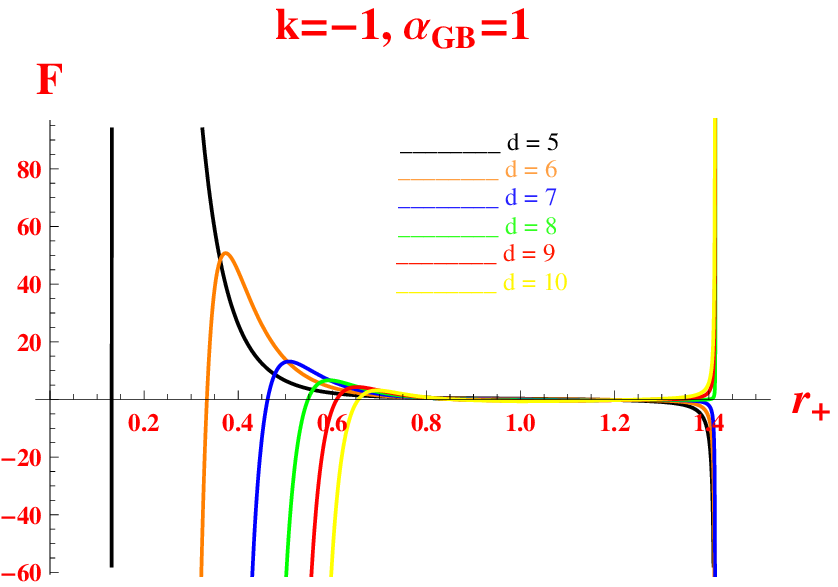}
\includegraphics[scale=.4]{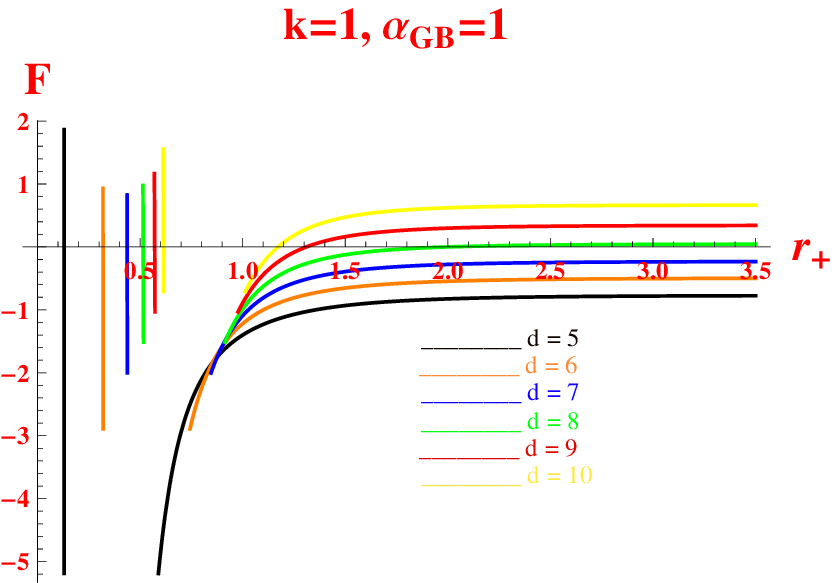}
\includegraphics[scale=.4]{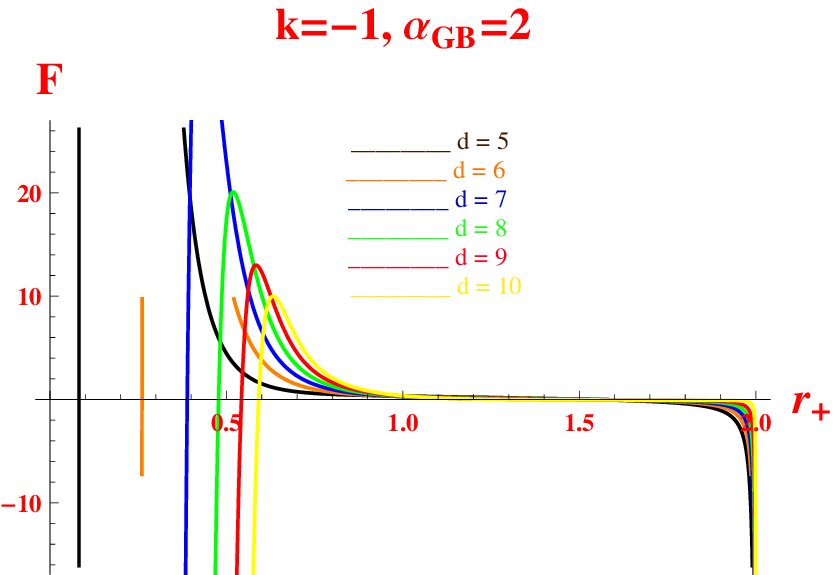}
\includegraphics[scale=.4]{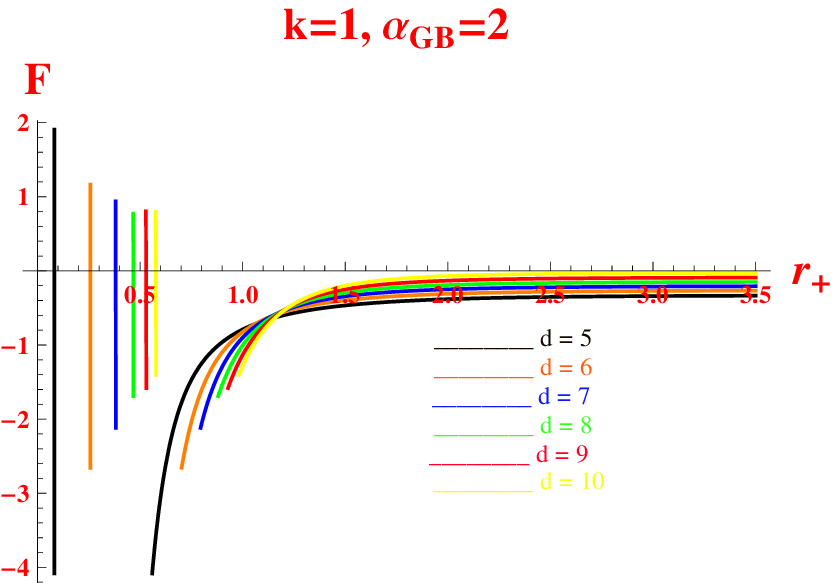}\\
Fig.-3.1a-3.1b represent the variations of $F$ with respect to horizon radius $ r_+ $ for $ k=-1, 1 $, $\alpha_{GB}=1$ and different dimensions $d$.\\
Fig.-3.2a-3.2b represent the variations of $F$ with respect to horizon radius $ r_+ $ for $ k=-1, 1 $, $\alpha_{GB}=2$ and different dimensions $d$.\\
\end{center}
\end{figure}
We have plotted the  variations of $F$ with respect to horizon radius $ r_+ $ for $ k=-1, 1 $, $\alpha_{GB}=1, 2$ and for different dimensions $d$ in Fig.-3.1a-3.1b and in Fig.-3.2a-3.2b. For Ricci flat $(k=0)$ BH horizon this thermodynamic property is undetermined and also for  hyperbolic topology of the BH horizon it would be imaginary. Hence the plot shown in Fig.-3.1a and Fig.-3.2a have no physical meaning. Therefore, only for spherical BH horizon of this type of BHs $F$ has physical significance.  
  
The internal energy $U$ is to be calculated from the familiar thermodynamic relation, computed as
 \begin{equation}\label{ah7_equn15}
U=F + T_+ S
\end{equation}
and which gives
$$U=\frac{r^{-d-7}}{640 (d-2) k^{3/2} \alpha ^{5/2} \pi  \left(2 k \alpha +r^2\right)}$$
   
$$\left(2 \sqrt{k} \sqrt{\alpha } \left(d r^{d+2} \left(2 k \alpha +r^2\right) \left(-12 (d-5) (d-2) k^2 \alpha ^2-10 (d-2) (d-1) k r^2 \alpha\right.\right.\right.$$
   
$$+15 r^4 ((d-3) d+8 \alpha  \Lambda +2)\bigg)+80 \pi  k \alpha ^2 r^{2 d} \left((d-2) k \left((d-5) k \alpha +(d-3) r^2\right)-2 r^4 \Lambda \right)$$
        
\begin{equation}\label{ah7_equn16}
 +40 \pi  k Q^2 r^8 \alpha ^2\big)+15 \sqrt{2} d r^{d+7} ((d-3) d+8 \alpha  \Lambda +2) \left(2 k \alpha +r^2\right) \tan ^{-1}\left(\frac{r}{\sqrt{2} \sqrt{k} \sqrt{\alpha }}\right)\bigg)
\end{equation}

\begin{figure}[h!]
\begin{center}
~~~~~~~Fig.-4.1a ~~~~~~~~~~Fig.-4.1b~~~~~~~~~~~~~~~Fig.-4.2a ~~~~~~~~~~Fig.-4.2b~~~~~~~~~~~~~~~\\
\includegraphics[scale=.4]{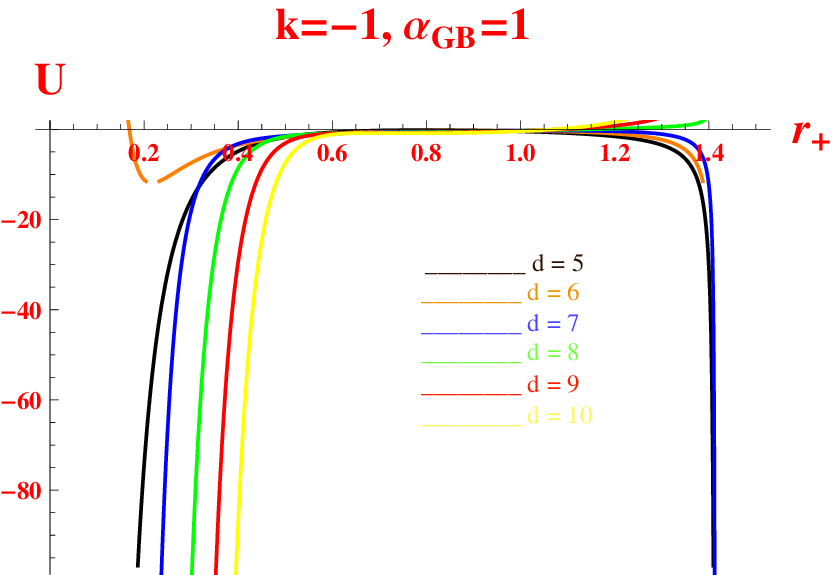}
\includegraphics[scale=.4]{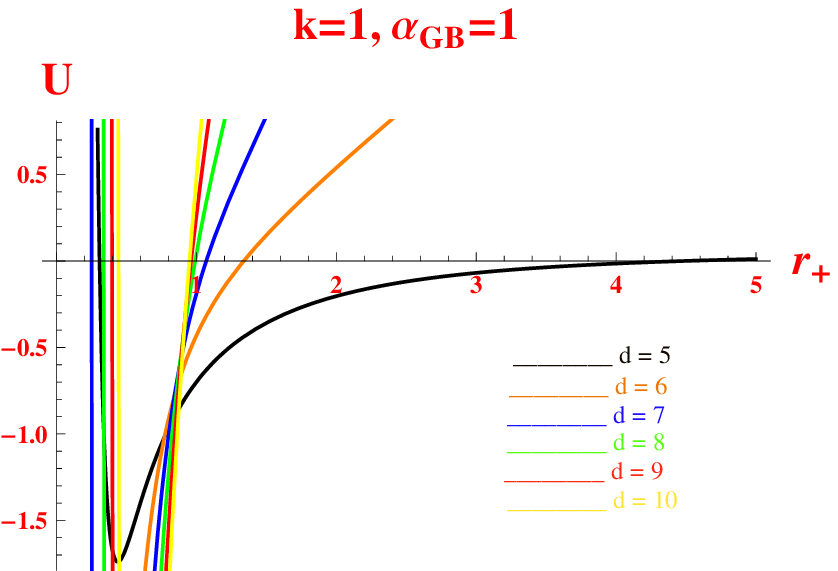}
\includegraphics[scale=.4]{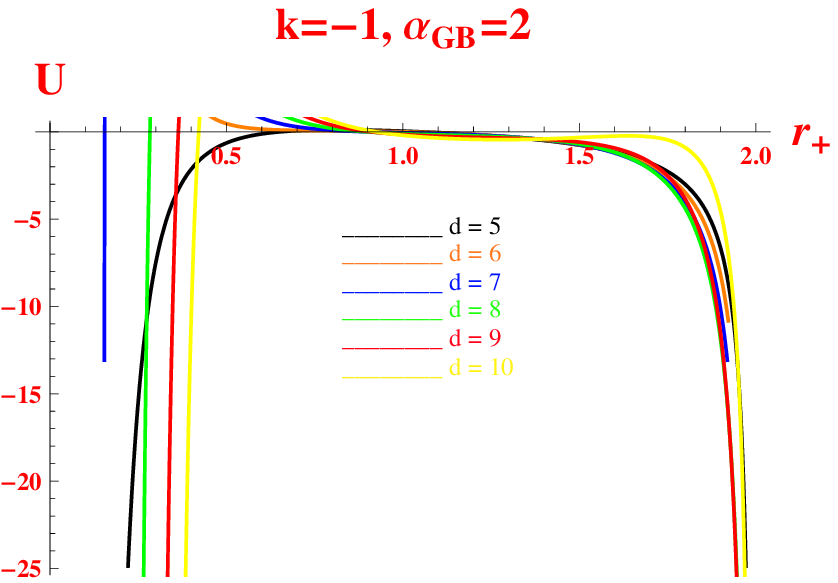}
\includegraphics[scale=.4]{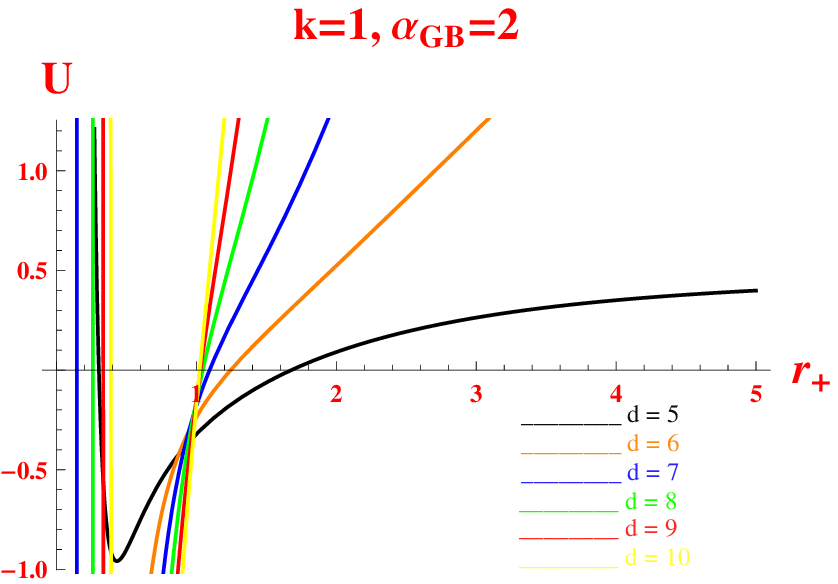}\\

Fig.-4.1a-4.1b represent the variations of $F$ with respect to horizon radius $ r_+ $ for $ k=-1, 1 $, $\alpha_{GB}=1$ and different dimensions $d$.\\
Fig.-4.2a-4.2b represent the variation of $F$ with respect to horizon radius $ r_+ $ for $ k=-1, 1 $, $\alpha_{GB}=2$ and different dimensions $d$.\\
\end{center} 
\end{figure}
We plot the internal energy in the Fig.-4.1a-4.1b (for $k=-1,1$ and $\alpha_{GB}=1$ respectively) and in the Fig.-4.2a-4.2b (for $k=-1,1$ and $\alpha_{GB}=2$ respectively). The nature of the curves for $k=1$ is completely agreed for the closed universe case. Here, the internal energy falls abruptly for a growing small BH. Then after reaching a local minima, the internal energy increases. The nature of the increasing part looks like a straight line. For $k=-1$ the internal energy is imaginary that means the curves shown in Fig.-4.1a and Fig.-4.2a have no physical significance. 

The thermodynamic volume $V$ of the BHs is expressed as 
\begin{equation}\label{ah7_equn17}
V=\frac{4}{3}\pi r_+^3
\end{equation}
and the modified pressure $P$ of the BH due to the thermal fluctuation can be derived from the relation,
\begin{equation}\label{ah7_equn18}
P=-\left(\frac{\partial F}{\partial V}\right)_T=-\left(\frac{\frac{\partial F}{\partial r_+}}{\frac{\partial V}{\partial r_+}}\right)_T
\end{equation}
which gives
$$P=\frac{r^{-2 (d+4)}}{64 (d-2) \pi ^2 \left(2 k \alpha +r^2\right)^2}\bigg[\bigg(2 r^{2 d} \bigg((d-2) k \left(10 (d-5) k^2 \alpha ^2+(13 d-53) k r^2 \alpha +5 (d-3) r^4\right)$$
$$-2 r^4   \Lambda  (2 k \alpha +3 r^2)\bigg)+Q^2 r^8 \left(2 (2 d-3) k \alpha +(2 d-1) r^2\right)\bigg)$$   
$$\log \left(\frac{r^{-3 d-4}  \left(2 r^{2 d} \left((d-2) k \left((d-5) k \alpha+(d-3) r^2\right)-2 r^4 \Lambda \right)+Q^2 r^8\right)^2}{64 (d-2)^2 \pi  \left(2 k  \alpha +r^2\right)^2}\right) $$   
$$+8 r^{2 d} \left((2-d) (d-1) k \left(2 (d-5) k^2 \alpha ^2+(3 d-10) k r^2 \alpha +(d-3) r^4\right)\right.$$        
\begin{equation}\label{ah7_equn19}
+2 r^4 \Lambda  \left(2 (d+1) k \alpha+d r^2\left)\left)\right.\right.\right.+Q^2 r^8 \left(6 (d-4) k \alpha +(3 d-8) r^2\right)\bigg]
\end{equation}
It is evident from $equation$ (\ref{ah7_equn19}) that when $r_+$ is increasing function of $\alpha_{GB}$ for small radius, pressure is decreasing function of $r_+$ for Ricci flat of the BH horizon, as expected. It is also noticed that for large event horizon radius the logarithmic correction does not play an important role on the pressure of the BH.

Another thermodynamic parameter `enthalpy' $H$ may be calculated from the relation given as:
\begin{equation}\label{ah7_equn20}
H=U+PV
\end{equation}
which gives
$$H=\frac{r^{-2 d-7}}{1920 (d-2) \pi  \left(2 k \alpha +r^2\right)^2}\bigg[40 r^2 \left(\bigg(2 r^{2 d} \left((d-2) k \left(10 (d-5) k^2 \alpha ^2+(13 d-53) k r^2 \alpha +5 (d-3) r^4\right)\right.\right.$$
$$-2r^4 \Lambda  \left(2 k \alpha +3 r^2\right)\bigg)+Q^2 r^8 \left(2 (2 d-3) k \alpha +(2 d-1) r^2\right)\bigg) $$
$$\log \bigg(\frac{r^{-3 d-4} \left(2 r^{2 d} \left((d-2) k \left((d-5) k \alpha +(d-3) r^2\right)
   -2 r^4 \Lambda \right)+Q^2 r^8\right)^2}{64 (d-2)^2 \pi \left(2 k \alpha +r^2\right)^2}\bigg)$$   
$$+8 r^{2 d} \left((2-d) (d-1) k \left(2 (d-5) k^2 \alpha ^2+(3 d-10) k r^2 \alpha\right.\right. $$
$$+(d-3) r^4\big)+2 r^4 \Lambda  \left(2 (d+1) k \alpha +d r^2\right)\big)+Q^2 r^8 \left(6 (d-4) k \alpha +(3 d-8) r^2\right)\big)$$
$$+\frac{3 r^d \left(2 k \alpha +r^2\right)}{{k^{3/2} \alpha ^{5/2}}}\bigg(2 \sqrt{k} \sqrt{\alpha } \left(d r^{d+2} \left(2 k \alpha +r^2\right) \left(-12 (d-5) (d-2) k^2 \alpha ^2-10 (d-2) (d-1) k r^2 \alpha\right.\right. $$
$$+15 r^4 ((d-3) d+8 \alpha  \Lambda +2)\big)+80 \pi  k \alpha ^2 r^{2 d} \left((d-2) k
 \left((d-5) k \alpha +(d-3) r^2\right)-2 r^4 \Lambda \right)$$
\begin{equation}\label{ah7_equn21}
+40 \pi  k Q^2 r^8 \alpha ^2\big)+15 \sqrt{2} d r^{d+7} ((d-3) d +8 \alpha  \Lambda +2) \left(2 k \alpha +r^2\right) \tan ^{-1}\left(\frac{r}{\sqrt{2} \sqrt{k} \sqrt{\alpha }}\right)\bigg)\bigg]
\end{equation} 
which is also a decreasing function of $\alpha_{GB}$ as well for the same.
\begin{figure}[h!]
\begin{center}
~~~~~~~Fig.-5.1a ~~~~~~~~~~Fig.-5.1b~~~~~~~~~~~~~~~Fig.-5.2a ~~~~~~~~~~Fig.-5.2b~~~~~~~~~~~~~~~\\
\includegraphics[scale=.4]{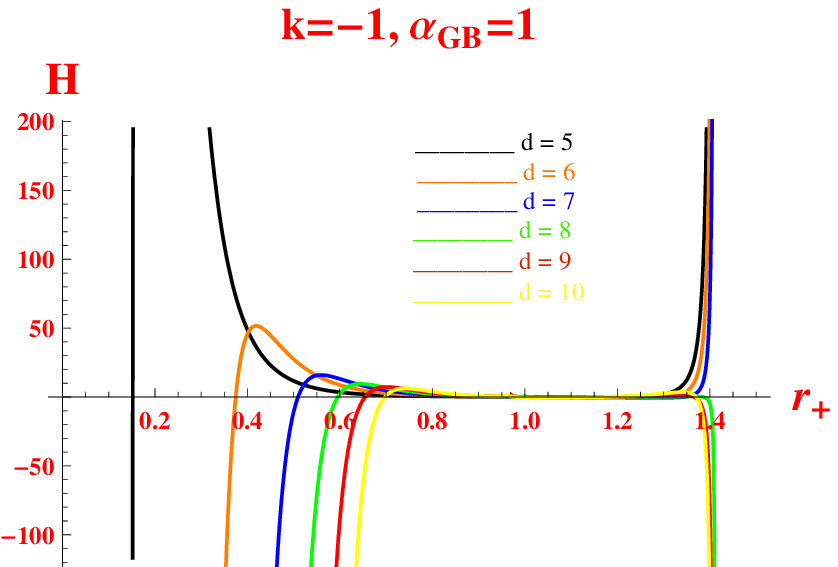}
\includegraphics[scale=.4]{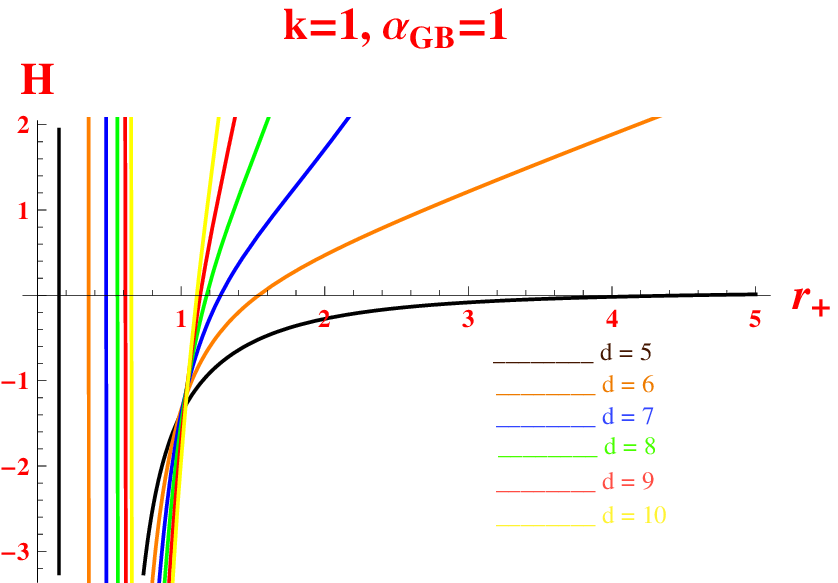}
\includegraphics[scale=.4]{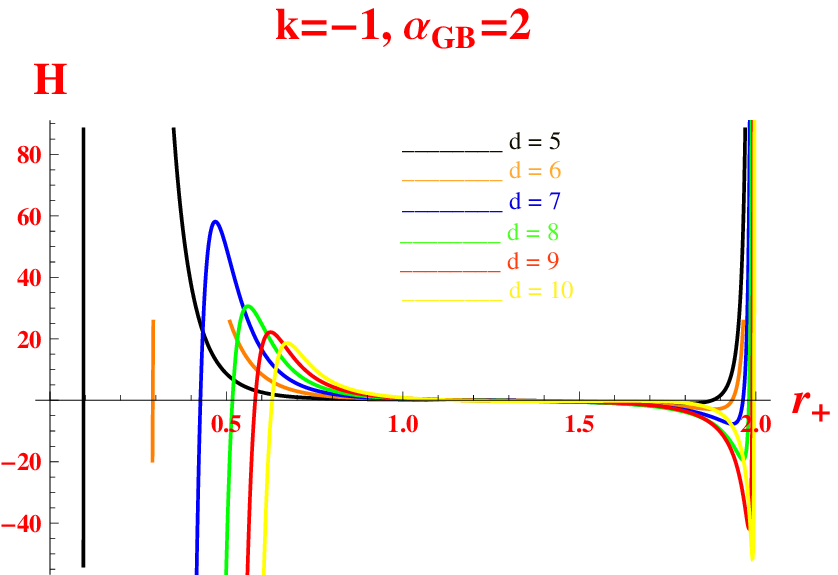}
\includegraphics[scale=.4]{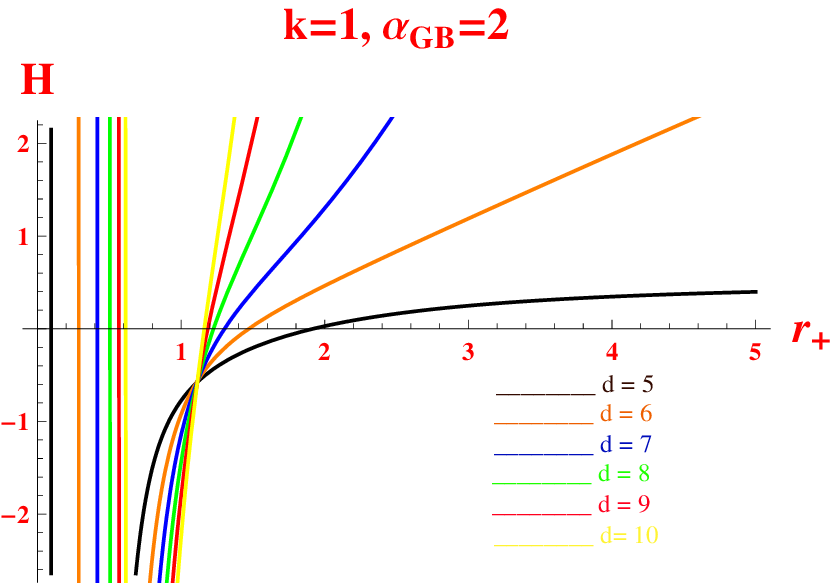}\\
Fig.-5.1a-5.1b represent the variation of $H$ with respect to horizon radius $ r_+ $ for $ k=-1, 1 $, $\alpha_{GB}=1$ and different dimensions $d$.\\
Fig.-5.2a-5.2b represent the variation of $H$ with respect to horizon radius $ r_+ $ for $ k=-1, 1 $, $\alpha_{GB}=2$ and different dimensions $d$.\\
\end{center}
\end{figure}
We plot the enthalpy $H$ in figures 5.1a-5.1b and 5.2a-5.2b (for $k=-1, 1$ and $\alpha_{GB}=1, 2$ respectively). Enthalpy seems to have the same nature as the internal energy when we consider closed (fig. 5.2b). On the other hand if an open universe is concerned, we see the enthalpy to decrease at first, then, after reaching a minima it increases rapidly with small increment of $r_+$ and for further increment of $r_+$, $H$ increases with a slow rate. The inclination of the curves increases with increase of the dimensions. For $k=-1$ the enthalpy is imaginary that implies the curves shown in Fig.-5.1a and Fig.-5.2a have no physical existences. 

The Gibbs free energy can be computed from the following thermodynamic relation as:
\begin{equation}\label{ah7_equn22}
G=H - T_+ S=F+PV,
\end{equation}
and which gives 
$$G=\frac{r^{-2 d-5}}{{1920 (d-2) \pi  \left(2 k \alpha +r^2\right)^2}}\bigg[\frac{45 \sqrt{2} d r^{2 d+5} ((d-3) d+8 \alpha  \Lambda +2) \left(2 k \alpha +r^2\right)^2 \tan ^{-1}\left(\frac{r}{\sqrt{2} \sqrt{k} \sqrt{\alpha }}\right)}{k^{3/2} \alpha ^{5/2}}$$
$$+\frac{1}{k \alpha ^2}\bigg\{40 k Q^2 r^8 \alpha ^2 \left(6 (d-4) k \alpha +(3 d-8) r^2\right)$$
$$+2 r^{2 d} \left(d^3 \left(2 k \alpha +r^2\right) \left(-232 k^3 \alpha ^3-256 k^2 r^2 \alpha ^2+60
   k r^4 \alpha +45 r^6\right)\right.$$
$$+d^2 \left(3568 k^4 \alpha ^4+4408 k^3 r^2 \alpha ^3+1032 k^2 r^4 \alpha ^2-450 k r^6 \alpha -135
   r^8\right)$$
$$+10 d \bigg(-688 k^4 \alpha ^4-744 k^3 r^2 \alpha ^3+8 k^2 r^4 \alpha ^2 (26 \alpha  \Lambda -25)+2 k r^6 \alpha (88 \alpha  \Lambda +15)$$
$$+9 r^8 (4 \alpha  \Lambda +1)\bigg)+320 k^2 \alpha ^2 \left(10 k^2 \alpha ^2+10 k r^2 \alpha +r^4 (2 \alpha  \Lambda +3)\right)$$ 
$$+80 k \alpha ^2 \bigg(2 r^{2 d} \left((d-2) k \left(8 (d-5) k^2 \alpha ^2+(11 d-43) k r^2 \alpha +4 (d-3) r^4\right)-2 r^4 \Lambda  \left(4 k \alpha +3 r^2\right)\right)$$
\begin{equation}\label{ah7_equn23}
 +Q^2 r^8 \left(2 d k \alpha +(d+1) r^2\right)\bigg)\log \left(\frac{r^{-3 d-4} \left(2 r^{2 d} \left((d-2) k \left((d-5) k \alpha +(d-3) r^2\right)-2 r^4 \Lambda \right)+Q^2 r^8\right)^2}{64 (d-2)^2 \pi  \left(2 k \alpha +r^2\right)^2}\right)\bigg\}\bigg]
\end{equation}
We find that Gibbs free energy is decreasing function of $\tilde{\alpha}$ like all other thermodynamic parameters which have already been studied for the same.
\begin{figure}[h!]
\begin{center}
~~~~~~~Fig.-6.1a ~~~~~~~~~~Fig.-6.1b~~~~~~~~~~~~~~~Fig.-6.2a ~~~~~~~~~~Fig.-6.2b~~~~~~~~~~~~~~~\\
\includegraphics[scale=.4]{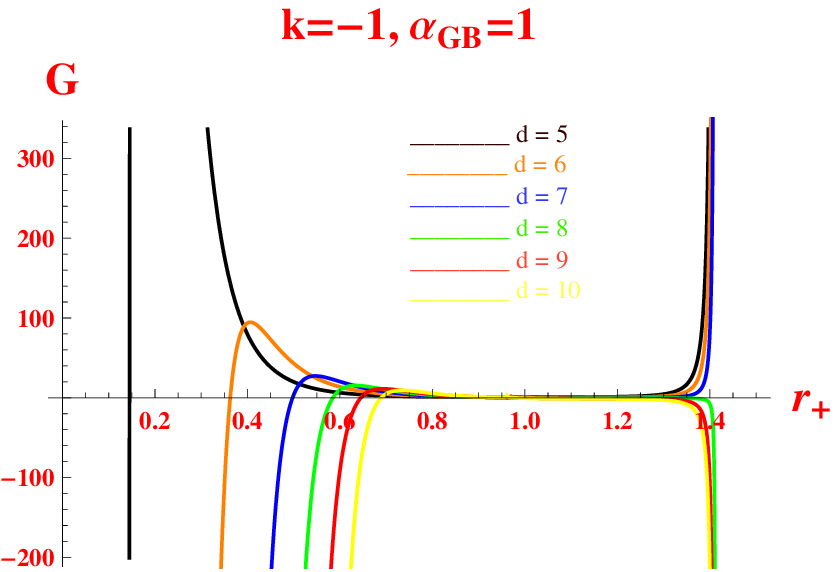}
\includegraphics[scale=.4]{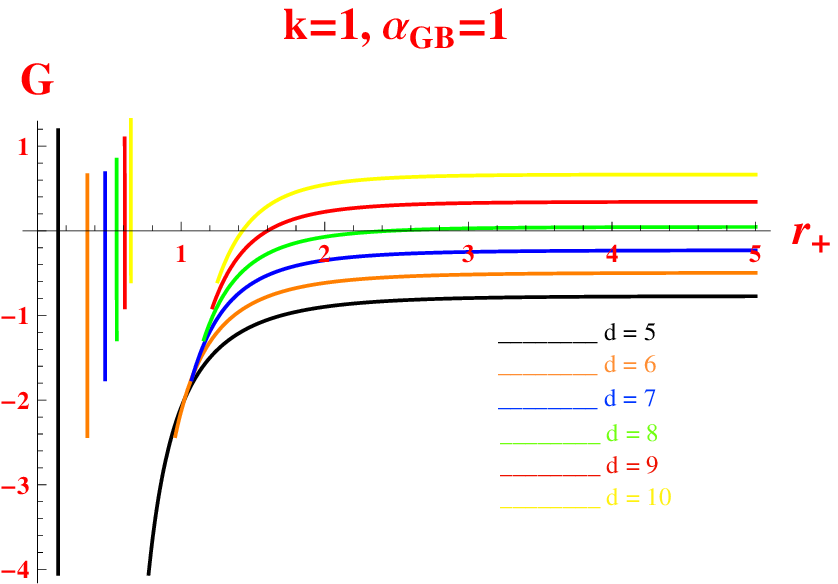}
\includegraphics[scale=.4]{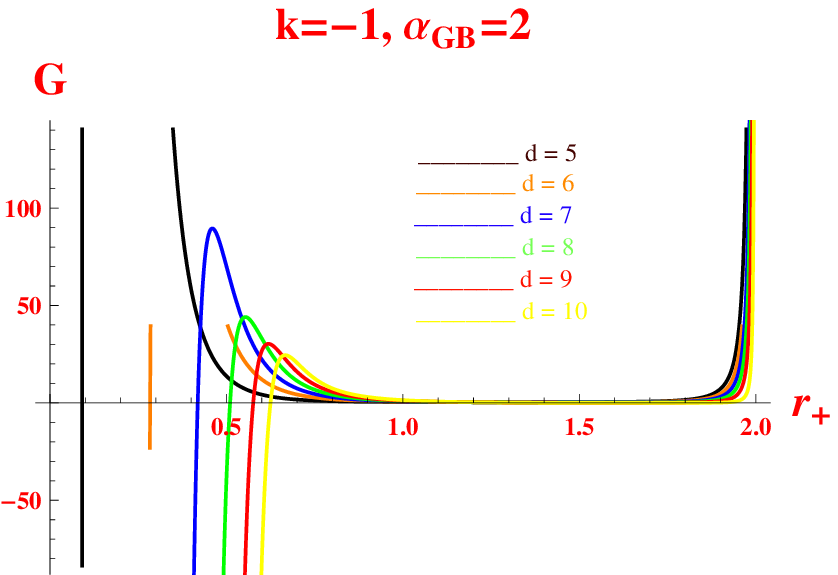}
\includegraphics[scale=.4]{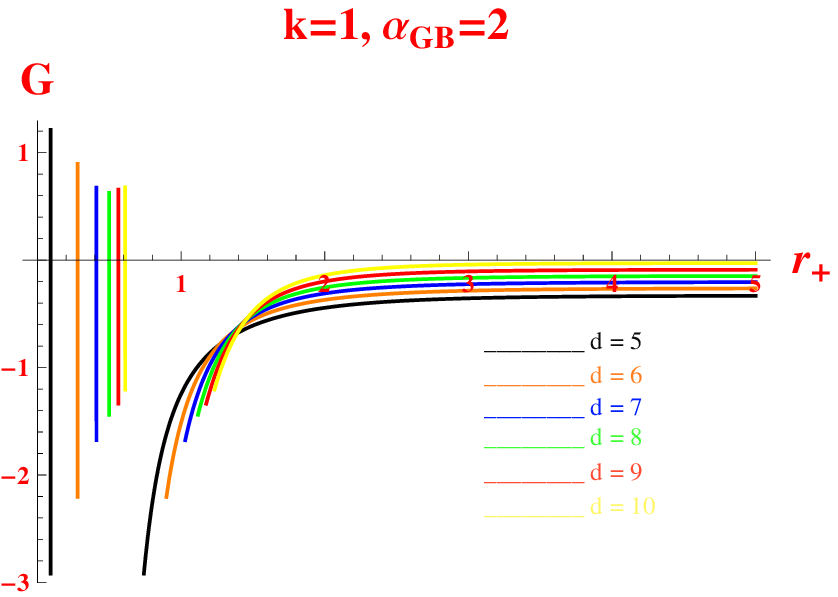}\\

Fig.-6.1a-6.1b represent the variation of $G$ with respect to horizon radius $ r_+ $ for $ k=-1, 1 $, $\alpha_{GB}=1$ and different dimensions $d$.\\
Fig.-6.2a-6.2b represent the variation of $G$ with respect to horizon radius $ r_+ $ for $ k=-1, 1 $, $\alpha_{GB}=2$ and different dimensions $d$.\\
\end{center}
\end{figure}
We have plotted the Gibb's free energy in figures 6.1a-6.1b and 6.2a-6.2b respectively (for $k=-1, 1$ and $\alpha_{GB}=1, 2$ respectively). Here we observe that for low values of $r_+$, $G$ decreases rapidly and reaches a local minimum and then for slight increment of $r_+$, $G$ increases rapidly and gets a saturated value.  For $k=-1$ the Gibb's free energy is imaginary which signifies that the curves shown in Fig.-6.1a and Fig.-6.2a have no physical significances. 
\section{$P$-$V$ Criticality Analysis}
The critical point is determined as the inflection point in the P-V diagram via the following equations:
$$\left.\frac{\partial P}{\partial V}\right| _{T = T_c}=0$$
and
\begin{equation}\label{ah7_equn24}
\left.\frac{\partial^2 P}{\partial V^2}\right| _{T = T_c}=0,
\end{equation}
In 5-dimensional space-time, at $\alpha_{GB}=1$ limit, one can obtain the points from both the equations of (\ref{ah7_equn24}) as:
\begin{equation}\label{ah7_equn25}
V_c =\frac{4}{3} \pi  \left(\frac{B}{\sqrt[3]{3} A}+\frac{A}{3^{2/3}}+4\right)^{3/2}~~.
\end{equation}
But the first condition of (\ref{ah7_equn24}) gives $T_c$ and $P_c$ as:
$$T_c=\frac{\frac{Q^2 \left(\frac{B}{3^{1/3} A}+\frac{A}{3^{2/3}}+8\right)}{\frac{B}{3^{1/3}A}+\frac{A}{3^{2/3}}+4}+\left(\frac{B}{3^{1/3} A}+\frac{A}{3^{2/3}}+4\right)^2+4 \left(\frac{B}{3^{1/3}A}+\frac{A}{3^{2/3}}+4\right)}{\pi  \left(\frac{B}{3^{1/3} A}+\frac{A}{3^{2/3}}+4\right)^{5/2}+16 \pi  \left(\frac{B}{3^{1/3} A}+\frac{A}{3^{2/3}}+4\right)^{3/2}+48 \pi  \sqrt{\frac{B}{3^{1/3} A}+\frac{A}{3^{2/3}}+4}}$$ and
$$P_c=\bigg[27 A \bigg\{\sqrt[3]{3} A^8+48 A^7+3^{2/3} A^6 \left(4 B+5 \left(Q^2+48\right)\right)+72 \sqrt[3]{3} A^5 \left(2 B+3 Q^2+16\right)+6 A^4 \bigg((5 B+384) Q^2$$

$$+3 B (B+80)\bigg)+72 3^{2/3} A^3 B \left(2 B+3 Q^2+16\right)+3 \sqrt[3]{3} A^2 B^2 \left(4 B+5 \left(Q^2+48\right)\right)+144 A B^3+3 3^{2/3} B^4\bigg\}\bigg]$$
\begin{equation}\label{ah7_equn26}
\bigg[8 \pi  \left\{A \left(\sqrt[3]{3} A+12\right)+3^{2/3} B\right\}^3 \left\{3^{2/3}A^4+72 \sqrt[3]{3} A^3+6 A^2 (B+192)+72 3^{2/3} A B+3 \sqrt[3]{3} B^2\right\}\bigg]^{-1}.
\end{equation}
Where as the second condition gives  
\begin{equation}\label{ah7_equn27}
T_c=\frac{\frac{7 Q^2 \left(\frac{B}{3^{1/3} A}+\frac{A}{3^{2/3}}+8\right)}{\frac{B}{3^{1/3} A}+\frac{A}{3^{2/3}}+4}+3\left(\frac{B}{3^{1/3} A}+\frac{A}{3^{2/3}}+4\right)^2+12 \left(\frac{B}{3^{1/3} A}+\frac{A}{3^{2/3}}+4\right)}{2 \pi \left(\frac{B}{3^{1/3} A}+\frac{A}{3^{2/3}}+4\right)^{5/2}+56 \pi \left(\frac{B}{3^{1/3}A}+\frac{A}{3^{2/3}}+4\right)^{3/2}+192 \pi  \sqrt{\frac{B}{3^{1/3} A}+\frac{A}{3^{2/3}}+4}}
\end{equation}
and 
$$P_c=\bigg[27 A \bigg\{\sqrt[3]{3} A^8+42 A^7+2 3^{2/3} A^6 \left(2 B+5 Q^2+72\right)+18 \sqrt[3]{3} A^5 \left(7 B+25 Q^2-16\right)+6 A^4\bigg(10 (B+84) Q^2$$
$$+3 (B (B+48)-384)\bigg)+18 3^{2/3} A^3 B \left(7 B+25 Q^2-16\right)+6 \sqrt[3]{3} A^2 B^2 \left(2 B+5Q^2+72\right)+126 A B^3+3 3^{2/3} B^4\bigg\}\bigg]$$
\begin{equation}\label{ah7_equn28}
\bigg[4 \pi  \left(A \left(\sqrt[3]{3} A+12\right)+3^{2/3} B\right)^3 \left(3^{2/3} A^4+108 \sqrt[3]{3} A^3+6 A^2 (B+336)+108 3^{2/3} A B+3 \sqrt[3]{3} B^2\right\}\bigg]^{-1},
\end{equation}
where $$A= \left\{\sqrt{3 Q^2 \left(Q^2+144\right)\left(125 Q^2-432\right)}+576-252 Q^2\right\}^{1/3}~~
and ~~ B=48-5 Q^2~~.$$
It is clear from the above expressions that they may not be satisfied simultaneously, which signifies that there is no critical point without thermal fluctuations. To have the critical points we should assume the effect of thermal fluctuations. 
\begin{figure}[h!]
\begin{center}
~~~~~~~~~~~~~~~~~~Fig.-7.1a ~~~~~~~~~~~~~~~Fig.-7.1b~~~~~~~~~~~~~~~~7.1c~~~~~~~~~~~~~~~~\\
\includegraphics[scale=.5]{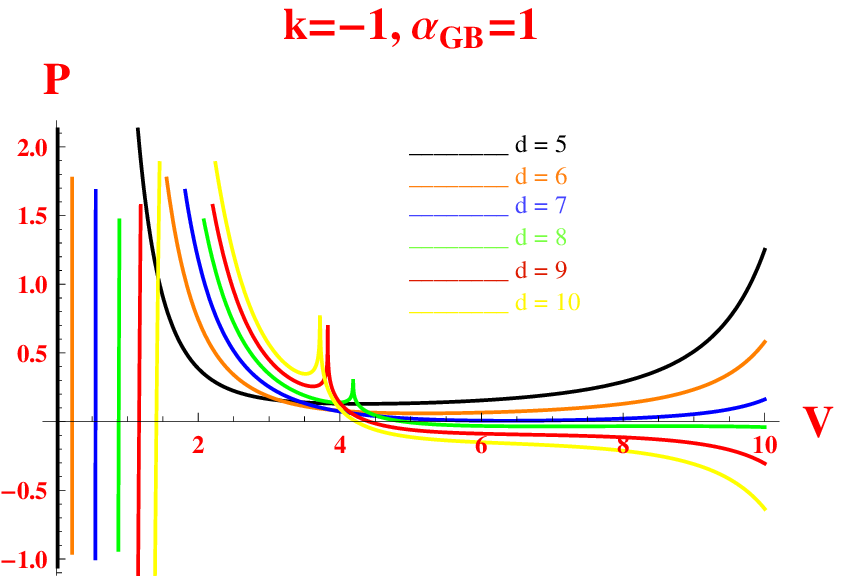}
\includegraphics[scale=.5]{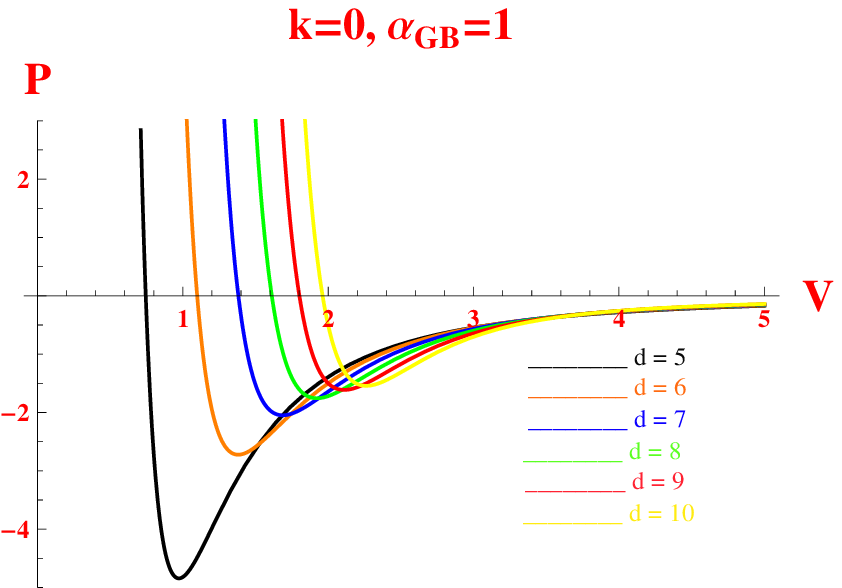}
\includegraphics[scale=.5]{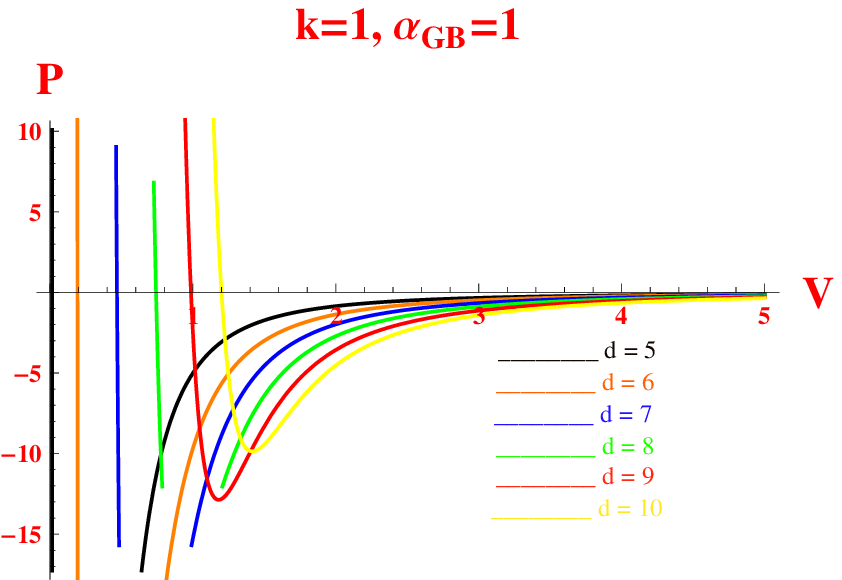}\\
~~~~~~~~~~~~~~~~~~Fig.-7.2a ~~~~~~~~~~~~~~~Fig.-7.2b~~~~~~~~~~~~~~~~7.2c~~~~~~~~~~~~~~~~\\
\includegraphics[scale=.5]{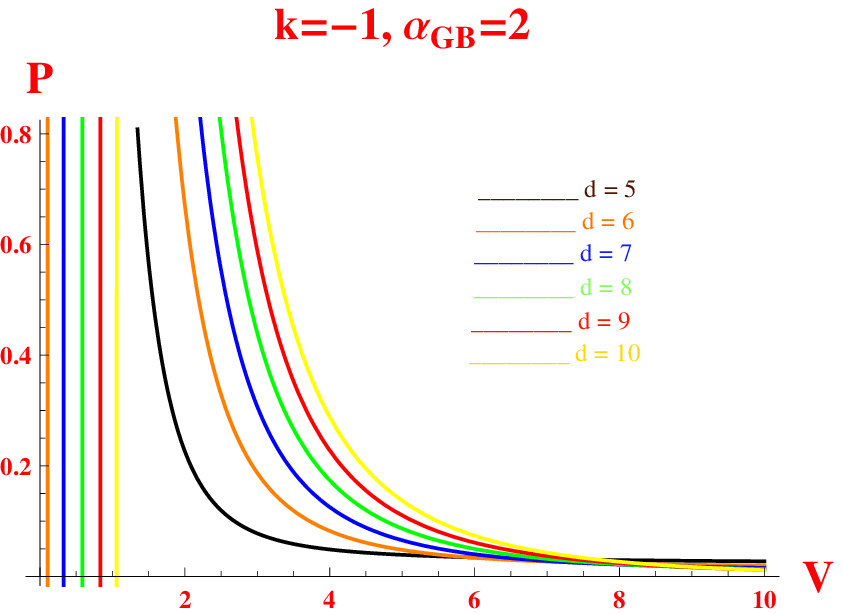}
\includegraphics[scale=.5]{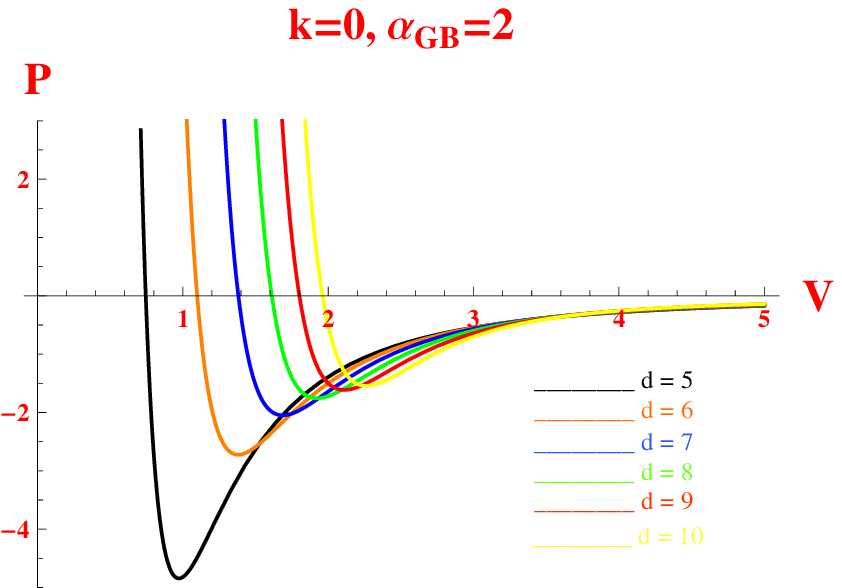}
\includegraphics[scale=.5]{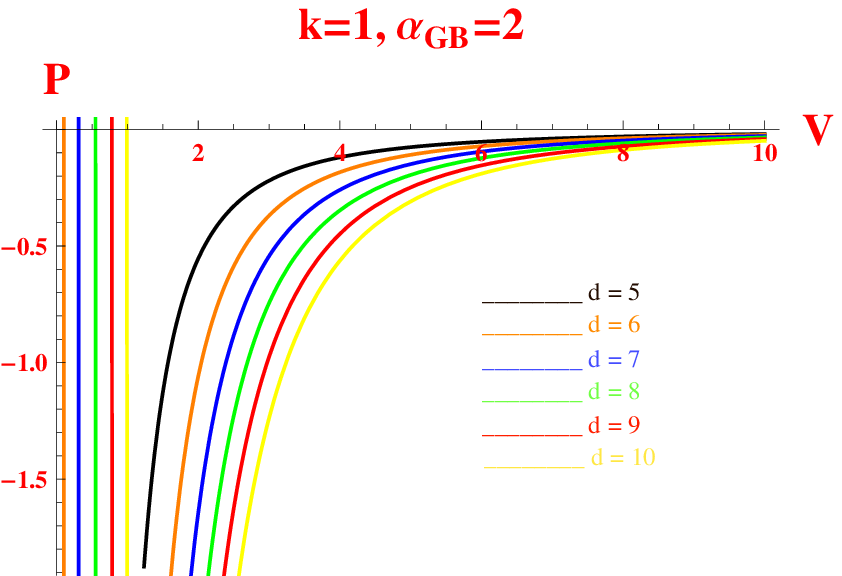}\\

Fig.-1.1a-1.1c represent the variation of $ P $ with respect to horizon radius $ V $ for $ k=-1, 0, 1 $, $\alpha_{GB}=1$ and different dimensions $d$. \\
Fig.-1.2a-1.2c represent the variation of $ P $ with respect to horizon radius $ V $ for $ k=-1, 0, 1 $, $\alpha_{GB}=2$ and different dimensions $d$. \\
\end{center} 
\end{figure}
Graphs for pressure vs volume are plotted in figures 7a-7c respectively ($k=-1,~0$ and $1$). For flat space, the curves typically resemble with the Boyle's law curves for ideal gas. For open universe, if dimension is low, some cusps are found. For closed universe, the $P$-$V$ curves are seen to have a local minima. 

Here we also compute the interesting between $P_c, T_c$ and $r_c\left(=\sqrt{\frac{B}{3^{1/3} A}+\frac{A}{3^{2/3}}+4}\right)$ as follows,
$$\frac{P_c r_c}{T_c}=\frac{3}{8}\bigg[\sqrt[3]{3} A^8+48 A^7+3^{2/3} A^6 \left(4 B+5 \left(Q^2+48\right)\right)+72 \sqrt[3]{3} A^5 \left(2 B+3 Q^2+16\right)+6 A^4\bigg\{5 B+384) Q^2$$
   
$$+3 B (B+80)\bigg\}+72 3^{2/3} A^3 B \left(2 B+3 Q^2+16\right)+3 \sqrt[3]{3} A^2 B^2 \left(4 B+5\left(Q^2+48\right)\right)+144 A B^3+3 3^{2/3} B^4\bigg]$$

$$\bigg[\left(A \left(\sqrt[3]{3} A+12\right)+3^{2/3} B\right) \bigg(A^6+16 3^{2/3}A^5+3 \sqrt[3]{3} A^4 \left(B+Q^2+80\right)$$
   
$$ +24 A^3 \left(4 B+3 Q^2+48\right)+3 3^{2/3} A^2 B \left(B+Q^2+80\right)+48 \sqrt[3]{3}A B^2+
3 B^3\bigg)\bigg]^{-1}$$
   
and    
   
$$\frac{P_c r_c}{T_c}=\frac{1}{2}\bigg[\sqrt[3]{3} A^8+42 A^7+2 3^{2/3} A^6 \left(2 B+5 Q^2+72\right)+18 \sqrt[3]{3} A^5 \left(7 B+25 Q^2-16\right)+6 A^4 \bigg(10 (B+84)Q^2$$

$$+3 (B (B+48)-384)\bigg)+18 3^{2/3} A^3 B \left(7 B+25 Q^2-16\right)+6 \sqrt[3]{3} A^2 B^2 \left(2 B+5 Q^2+72\right)+126 A B^3+33^{2/3} B^4\bigg]$$
   
$$ \bigg[\left(A \left(\sqrt[3]{3} A+12\right)+3^{2/3} B\right) \bigg(A^6+16 3^{2/3} A^5+\sqrt[3]{3} A^4 \left(3 B+7 Q^2+240\right)$$
\begin{equation}\label{ah7_equn29}
+24 A^3 \left(4 B+7 Q^2+48\right)+3^{2/3} A^2 B \left(3 B+7 Q^2+240\right)+48 \sqrt[3]{3} A B^2+
3 B^3\bigg)\bigg]^{-1}~~.
\end{equation}
\section{Thermal Stability Analysis}  
Now we analyse  the stability of the BH by employing the specific heat which is given by,
\begin{equation}\label{ah7_equn30}
C=T\left(\frac{dS}{dT}\right)=T\left(\frac{\frac{dS}{dr_+}}{\frac{dT}{dr_+}}\right) ,
\end{equation}
and examine it in different regime. One can obtain the specific heat of the BH utilizing $equation$ (\ref{ah7_equn27}) as:
$$C=-2 r_+^2 \bigg[2 r^2 \bigg\{Q^2 r_+^6 \left(r_+^2 (d+2 \pi  k \tilde{\alpha} -3)+2 (d-4) k \tilde{\alpha} +\pi r_+^4\right)$$
$$+2 r_+^{2 d} \bigg((d-2) k \bigg(\pi  \left(2 k\tilde{\alpha} +r_+^2\right) \left(d-5) k \tilde{\alpha} +(d-3) r_+^2\right)$$

$$+(1-d) k \tilde{\alpha} \bigg)+r_+^2 \Lambda  \left(r_+^2 (1-2 \pi  k \tilde{\alpha} )+4 k \tilde{\alpha}-\pi r_+^4\right)\bigg)\bigg\}\bigg]$$

$$\bigg[2 r_+^{2 d} \bigg((d-2) k \left(2 (d-5) k^2 \tilde{\alpha} ^2+(d-9) k r_+^2 \tilde{\alpha} +(d-3) r_+^4\right)$$

\begin{equation}\label{ah7_equn31}
+r_+^4 \Lambda \left(6 k \tilde{\alpha} +r_+^2\right)\bigg)+Q^2 r_+^8 \left(2 (2 d-7) k \tilde{\alpha} +(2 d-5) r_+^2\right)\bigg]^{-1}
\end{equation}

\begin{figure}[h!]
\begin{center}
~~~~~~~Fig.-8a ~~~~~~~~~~~~~~Fig.-8b~~~~~~~~~~~~~Fig.-8c~~~~~~~~~~~~~\\
\includegraphics[scale=.5]{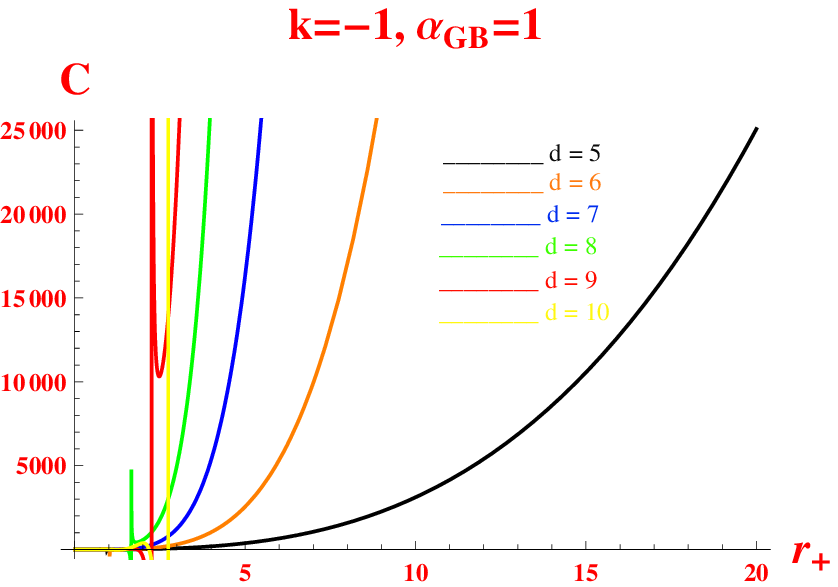}
\includegraphics[scale=.5]{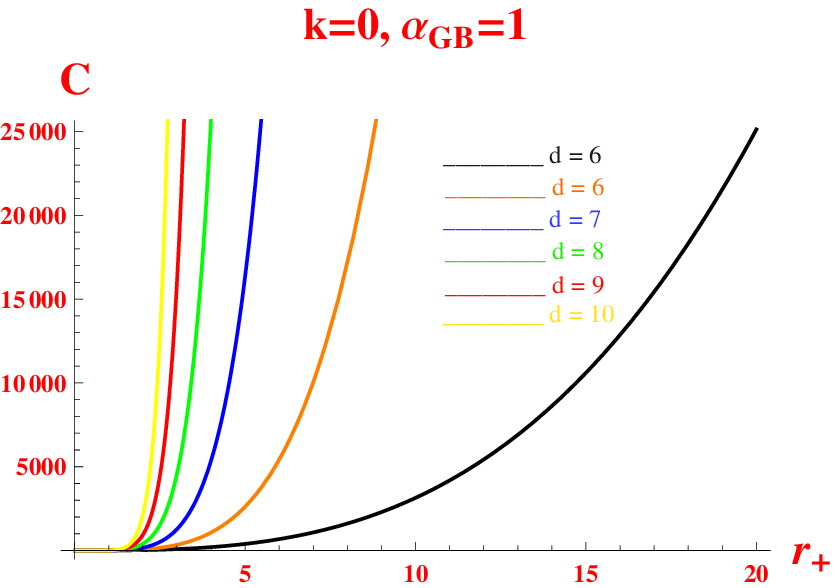}
\includegraphics[scale=.5]{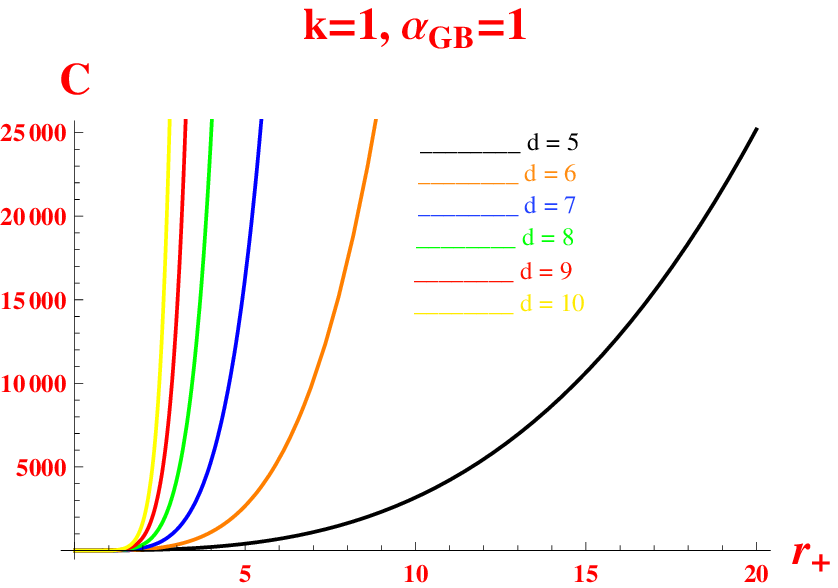}\\
Fig.-8a-8c represent the variation of $C$ with respect to horizon radius $ r_+ $ for $ k=-1, 0, 1 $ and different dimensions $d$.
\end{center} 
\end{figure}
Figures 8a-8c depict the variations of specific heat for open space, mainly two first order phase transitions take place. Firstly, from unstable small to stable intermediate mass BHs and then to unstable large massive BHs. Flat space allow first order phase transition from unstable to stable phases. In closed space, two second order phase transitions take place(unstable to stable to unstable again).
\section{Geometrothermodynamics}
In this section, we compute the Weinhold metric for the thermodynamic space of the concerned type of BHs. After huge calculation we obtain it into the form, shown in equation (\ref{ah7_equn32}) of Appendix-I. 
The corresponding Ruppeiner metric can be easily obtained by dividing it by the BHs' temperature at event horizon, $T_+$. We calculate the Ricci scalar for this metric. We plot this Ricci scalar with respect to $r_+$ and $Q$ in the Fig.-9.1a-9.1c and Fig.-9.2a-9.2c for $k=-1, 0, 1$ and $\alpha_{GB}=1, 2$, i.e., for the spherical, Ricci flat and hyperbolic topology of the BH horizon, respectively.
\begin{figure}[h!]
\begin{center}
~~~~~~~Fig.-9.1a ~~~~~~~~~~~~~~~~~~~~F-g.-9.1b~~~~~~~~~~~~~~~~Fig.-9.1c~~~~~~~~~~~~~~~~~~~~\\
\includegraphics[scale=.5]{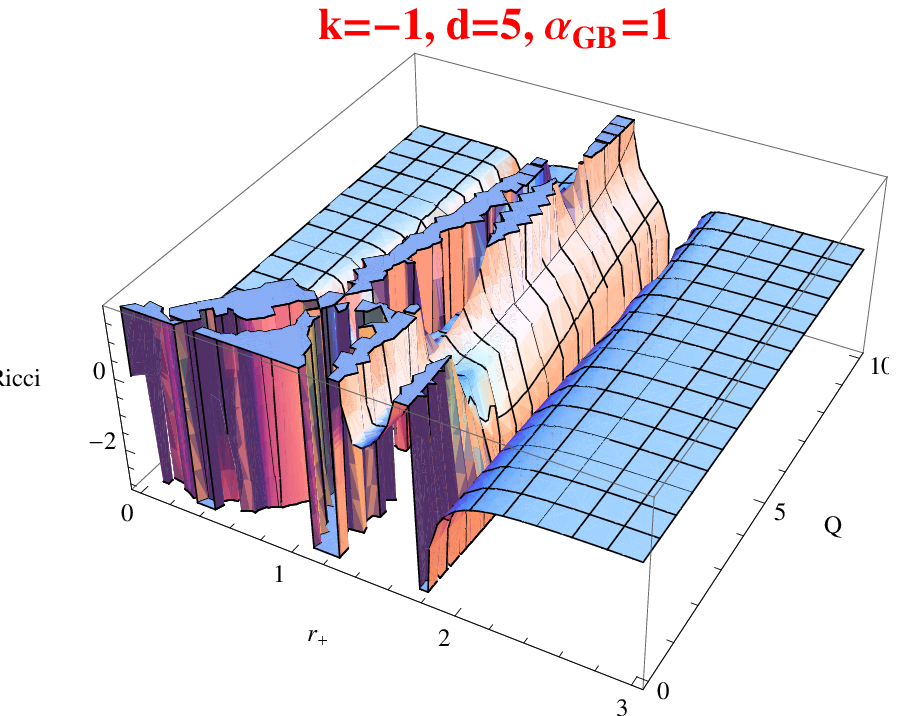}
\includegraphics[scale=.5]{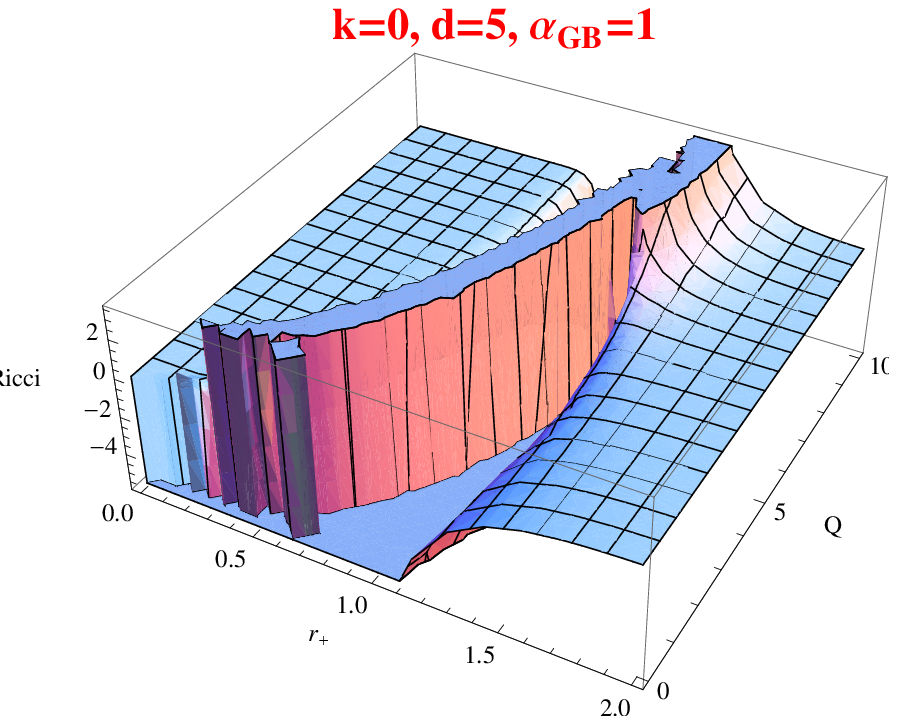}
\includegraphics[scale=.5]{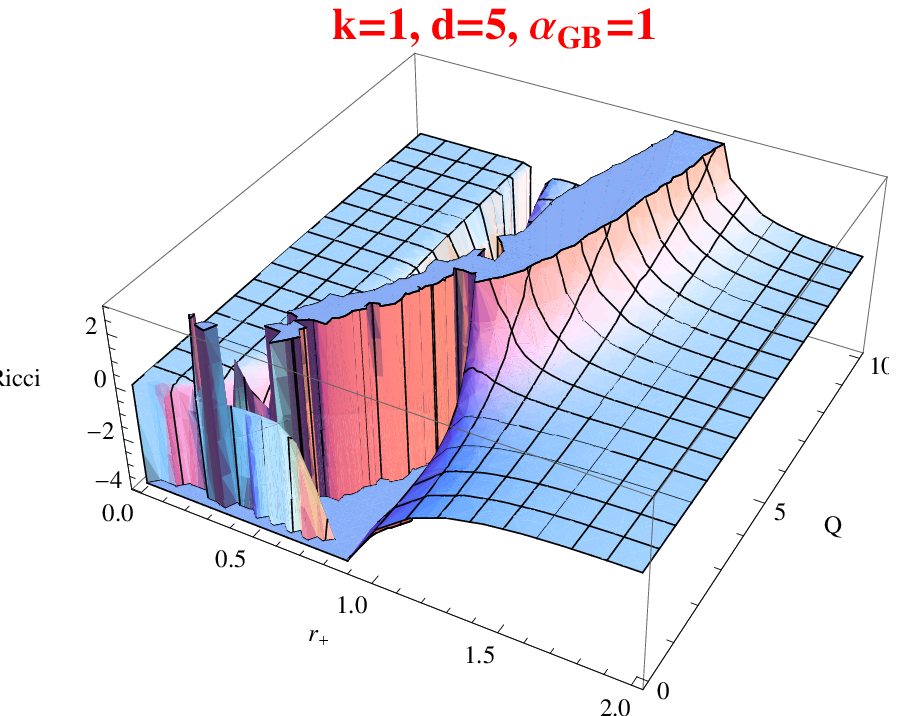}~~\\
~~~~~~~Fig.-9.2a ~~~~~~~~~~~~~~~~~~~~F-g.-9.2b~~~~~~~~~~~~~~~~Fig.-9.2c~~~~~~~~~~~~~~~~~~~~\\
\includegraphics[scale=.5]{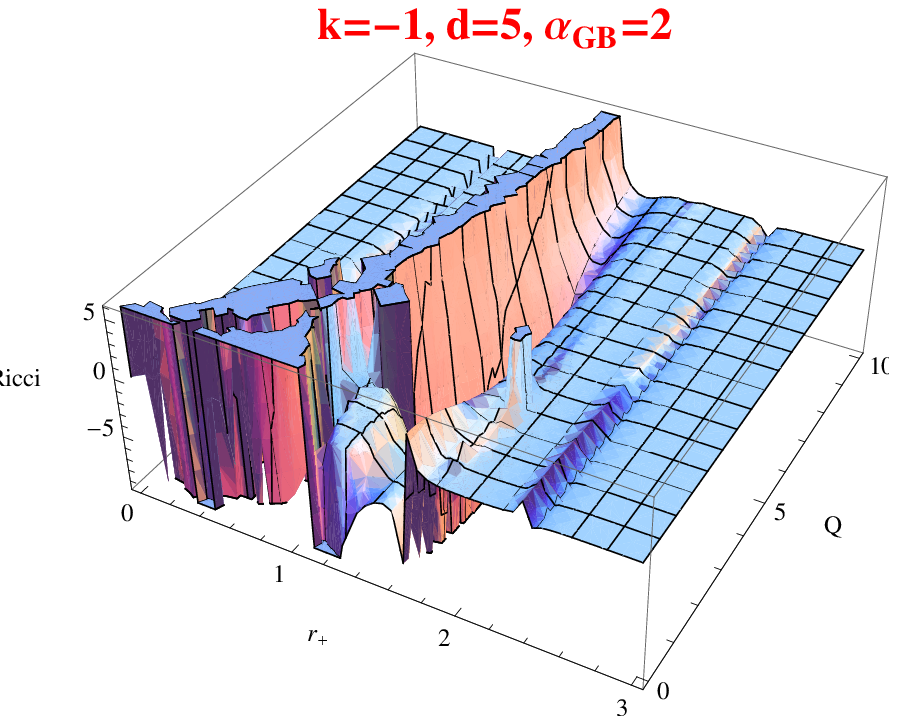}
\includegraphics[scale=.5]{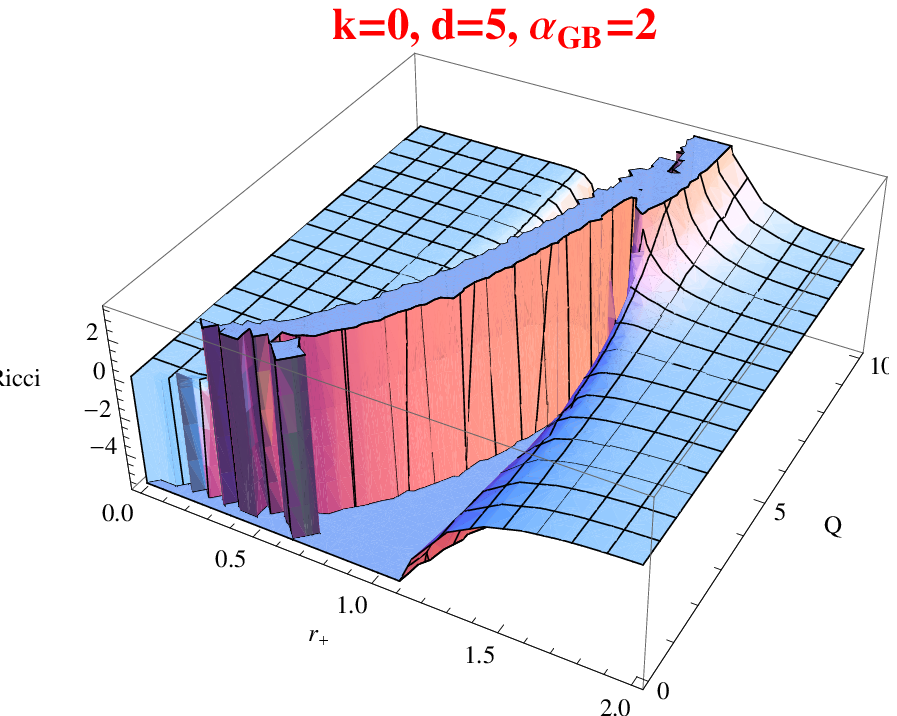}
\includegraphics[scale=.5]{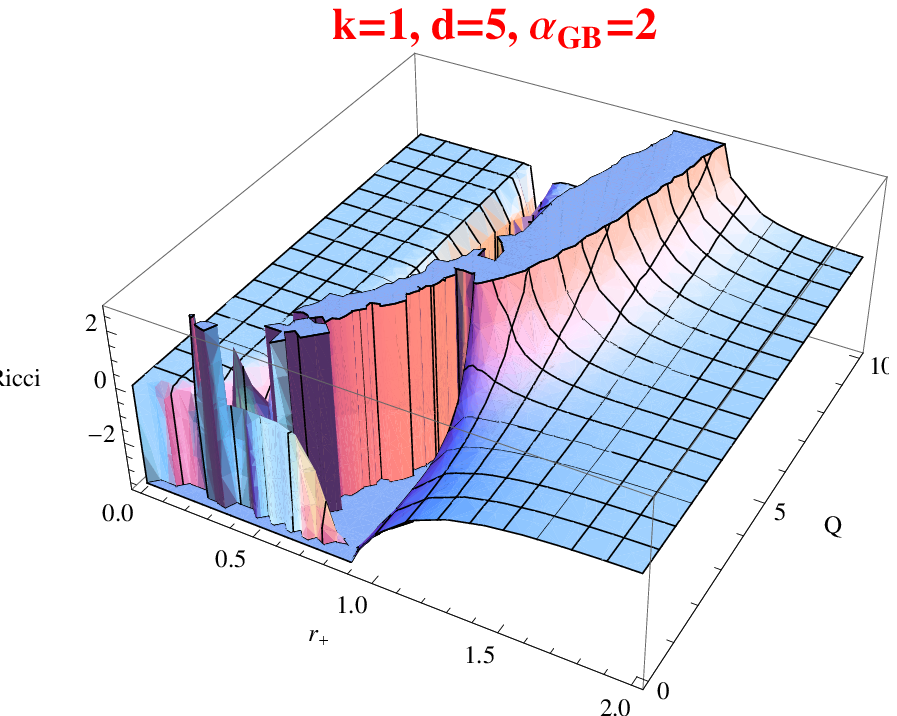}~~\\

Fig.-9.1a-9.1c represent the variation of $ Ricci~Scalar $ of Ruppeiner metric with respect to horizon radius $ r_+ $ and electric charge $Q$ for $k=-1, 0,  1$ and $\alpha_{GB}=1$.\\
Fig.-9.2a-9.2c represent the variation of $ Ricci~Scalar $ of Ruppeiner metric with respect to horizon radius $ r_+ $ and electric charge $Q$ for $k=-1, 0,  1$ and $\alpha_{GB}=2$.\\

\end{center} 
\end{figure}
The graphs of Ricci scalars(Fig.-9.1a-9.1c and Fig.-9.2a-9.2c) depict several divergences are possible for different values of $r_+$ and charge $Q$.
\section{Conclusions}
We have seen that higher dimensional background increases the possibilities to have a naked singularity rather than a BH spacetime wrapped by an event horizon \cite{Rudra1, Rudra2}. That is, the BHs are less stable in higher dimensions. Besides it has been seen that the Gauss-Bonnet coupling parameter makes the embedded BH more unstable \cite{Biswas1}. In this paper, our motivation was to use the first order entropy correction as a catalyst to increase the joint effect of Gauss-Bonnet coupling parameter and higher dimension. We see the temperature of BHs in an open Gauss-Bonnet universe is unphysical but it is decreasing when we take flat or closed universe. Free energy for closed universe changes sign twice and the internal energy shows extremal values. Enthalpy on the other hand has two branches. Firstly, it is increasing and then is decreasing. Gibb's free energy has cuspidal type double points. We have analysed $P$ vs $V$ curves and have shown the deviate from Boyle's law nature whenever open or closed universe is considered. Study of specific heat shows second order phase transition for closed universe and first order phase transition for flat or open universe. Ricci scalars have diverging lines. All these thermodynamic entities, as a whole, show that the BHs are less stable with first order correction of entropy, particularly, when higher dimensions are considered.

\vspace{.1 in}
{\bf Acknowledgment:}
This research is supported by the project grant of Goverment of West Bengal, Department of Higher Education, Science and Technology and Biotechnology (File no:- $ST/P/S\&T/16G$-$19/2017$). AH thanks Department of Mathematics, The University of Burdwan for providing research facilities. RB thanks IUCAA, Pune, India for providing Visiting Associateship.

RB dedicates this article to Prof. Subenoy Chkaraborty, Department of Mathematics, Jadavpur University, India to tribute him on his $60^{th}$ birth year.


\section{Appendix-I}
The Weinhold Metric is expressed as: 
\begin{equation}\label{ah7_equn32}
ds^2_W=\frac{\partial^2 M}{\partial S^2} dS^2+2\frac{\partial^2 M}{\partial S \partial Q} dMdQ+\frac{\partial^2 M}{\partial Q^2} dQ^2,
\end{equation} where

$$\frac{\partial^2 M}{\partial S^2}=-\left[(d-2) V r^{-d-3} \left(2 k \alpha +r^2\right)^2 \big(Q^2 r^8-2 r^{2 d} \big((d-6) (d-5) k^2 \alpha\right. $$

$$+(d-4) (d-3) k r^2-2 r^4 \Lambda \big)\big) \left(2 r^{2 d} \left((d-2) k \left((d-5) k \alpha +(d-3) r^2\right)-r^4 \Lambda \big)+Q^2 r^8\right)^2\right]$$
   
$$\bigg[32 \pi \big(2 Q^4 r^{18} \left(-4 (d-4) k^2 \alpha ^2+2 (9-2 d) k r^2 \alpha +(3-d) r^4\right)+8 r^{4 d+4} \big((d-2)^2 (d-1) k^3 \alpha  \big(-2 (d-5) k^2 \alpha ^2$$
   
$$+(3 d-11) k r^2 \alpha +3 (d-3) r^4\big)+(d-2) k r^2 \Lambda  \left(24 (d-5) k^3 \alpha ^3+8 (2 d-11) k^2 r^2 \alpha ^2+(d-3) r^6-16 k r^4 \alpha \right)$$

$$+r^6 \Lambda ^2 \left(8 k^2 \alpha ^2+10 k r^2 \alpha +r^4\right)\big)-4 Q^2 r^{2 (d+5)} \big(4 (d-5) (d-4) (d-2) (2 d-7) k^4 \alpha ^3$$
   
$$+8 (d-3) (d-2) (d (2 d-15)+29) k^3 r^2 \alpha ^2+k^2 r^4 \alpha  (d (d (d (10  d-113)-8 \alpha  \Lambda +471)+28 \alpha  \Lambda -864)-8 \alpha  \Lambda +588)$$
   
$$+k r^6 \left((d-3)^2 (d-2) (2 d-5)-4 (d (2 d-7)+1) \alpha
    \Lambda \right)+((7-2 d) d-4) r^8 \Lambda \big)$$
    
    $$+\pi  (d-3) (d-2) Q^4 r^{d+16} \left(2 k \alpha +r^2\right)^2+4 \pi  (d-3) (d-2) Q^2$$
    
$$
   r^{3 d+8} \left(2 k \alpha +r^2\right)^2 \left((d-2) k \left((d-5) k \alpha +(d-3) r^2\right)-r^4 \Lambda \right)$$

$$+4 \pi  (d-3) (d-2) r^{5 d} \left(2 k \alpha +r^2\right)^2
 \left\{(d-2) k \left((d-5) k \alpha +(d-3) r^2\right)-r^4 \Lambda \right\}^2\bigg]^{-1}$$

$$\frac{\partial^2 M}{\partial S \partial Q}=\bigg[Q V r^{5-d} \left(2 k \alpha +r^2\right) \left(4 Q^2 r^{2 d+8} \left(-2 (d-2) (d-1) k^2 r^4 \alpha +\pi  (d-2) r^d \left(2 k \alpha+r^2\right)\right.\right.$$ 

$$ \left((d-2) k \left((d-5) k \alpha +(d-3) r^2\right)-r^4 \Lambda \right)+8 k r^6 \alpha  \Lambda +2 r^8 \Lambda \left)\right.$$ 

$$+4 r^{4 d} \left(2 r^2 \left(-2 (d-5)^2 (d-4) (d-2)^2 k^5 \alpha ^3+(5-d) (d-2)^2 (d (5 d-34)+61) k^4 r^2 \alpha ^2\right.\right.$$

$$-2 (d-2) k^3 r^4 \alpha (2 (d-3) (d-2) ((d-7) d+13)-(d-5) (3 d-8) \alpha  \Lambda )$$ 

$$+(d-2) k^2 r^6 \left((d (9 d-53)+80) \alpha  \Lambda -(d-3)^3 (d-2)\right)+k  r^8 \Lambda  (d (d (3 d-22)-4 \alpha  \Lambda +53)$$

$$+4 \alpha  \Lambda -42)+(3-2 d) r^{10} \Lambda ^2\big)+\pi  (d-2) r^d \left(2 k \alpha +r^2\right) \left((d-2) k \left((d-5) k \alpha+(d-3) r^2\right)-r^4 \Lambda \right)^2\big)$$
   
$$+Q^4 r^{16} \left(4 (d-4) k r^2 \alpha +\pi  (d-2) r^d \left(2 k \alpha +r^2\right)+2 (d-3) r^4\right)\big)\bigg]$$
   
$$\bigg[16 \pi  \left(Q^2 r^8 \left(4 (d-4) k r^2 \alpha +\pi 
   (d-2) r^d \left(2 k \alpha +r^2\right)+2 (d-3) r^4\right)\right.$$
   
$$+4 r^{2 d+4} \left(k \alpha  \left((2-d) (d-1) k+4 r^2 \Lambda \right)+r^4
   \Lambda \right)$$   
   
$$ +2 \pi  (d-2) r^{3 d} \left(2 k \alpha +r^2\right) \left\{(d-2) k \left((d-5) k \alpha +(d-3) r^2\right)-r^4 \Lambda\big)\right\}^2\bigg]^{-1}$$   
   and 

$$\frac{\partial^2 M}{\partial Q^2}=\frac{V r_+^{3-d}}{48 \pi -16 d \pi }$$

 \end{document}